\documentstyle[12pt,psfig]{article}

\begin{document}

\newcommand{\gtrsim}{ \mathop{}_{\textstyle \sim}^{\textstyle >} }
\newcommand{\lesssim}{ \mathop{}_{\textstyle \sim}^{\textstyle <} }

\newcommand{\rem}[1]{{\bf #1}}

\renewcommand{\thefootnote}{\fnsymbol{footnote}}
\setcounter{footnote}{0}
\begin{titlepage}

\def\thefootnote{\fnsymbol{footnote}}

\begin{center}

\hfill TU-619\\
\hfill hep-ph/0104263\\
\hfill April, 2001\\

\vskip .75in

{\Large \bf 
CP Violation in Kaon System in Supersymmetric SU(5) Model
with Seesaw-Induced Neutrino Masses
}

\vskip .75in

{\large
Naoyuki Akama, Yuichiro Kiyo, Shinji Komine and Takeo Moroi
}

\vskip 0.25in

{\em
Department of Physics, Tohoku University,
Sendai 980-8578, JAPAN}

\end{center}

\vskip .5in

\begin{abstract}

  CP violations in the kaon system are studied in supersymmetric SU(5)
  model with right-handed neutrinos.  We pay a special attention to
  the renormalization group effect on the off-diagonal elements of the
  squark mass matrices.  In particular, if the Yukawa couplings and
  mixings in the neutrino sector are sizable, off-diagonal elements of
  the right-handed down-type squark mass matrix are generated, which affect CP
  and flavor violations in decay processes of the kaon.  We calculate
  supersymmetric contributions to $\epsilon$ (as well as $\Delta
  m_K$), $Br(K_L\rightarrow\pi^0\nu\bar{\nu})$, and
  $\epsilon'/\epsilon$ in this framework.  We will see that the
  supersymmetric contribution to the $\epsilon$ parameter can be as
  large as (and in some case, larger than) the experimentally measured
  value.  We also discuss its implication to future tests of the
  unitarity triangle of the Kobayashi-Maskawa matrix.

\end{abstract}
\end{titlepage}

\renewcommand{\theequation}{\thesection.\arabic{equation}}
\renewcommand{\thepage}{\arabic{page}}
\setcounter{page}{1}
\renewcommand{\thefootnote}{\#\arabic{footnote}}
\setcounter{footnote}{0}

\section{Introduction}
\label{sec:intro}
\setcounter{equation}{0}

One of the most important issues in particle physics today is to
understand the origin of CP violations.  Indeed, many efforts have
been made to measure CP violations in various processes.  So far, for
the $K$ system, non-vanishing values of the $\epsilon$ and $\epsilon'$
parameters have been observed.  In addition, CP violation in the
$B^0\rightarrow\psi K^0$ process, i.e., the angle
$\phi_1$,\footnote{The angle $\phi_1$ is also called the angle
$\beta$.  In this paper, however, we do not use this notation since we
use the angle $\beta$ to parameterize the vacuum expectation values of
two Higgs bosons.} is now being measured by the on-going
$B$-factories, and even at the present stage non-vanishing CP
violation in this process is reported \cite{BELLE,BABAR}.

In the framework of standard model (SM), the most well-known mechanism
to explain these CP violations is to introduce the Kobayashi-Maskawa
(KM) matrix \cite{PTP49-652} which contains one physical complex
phase for the three family case.  That is, using the Wolfenstein
parameterization \cite{wolfenstein51}:
\begin{eqnarray}
    V_{\rm KM} \simeq \left(
      \begin{array}{ccc}
          1-\lambda^2/2 & \lambda & A \lambda^3 (\rho-i\eta) \\
          -\lambda & 1-\lambda^2/2 & A \lambda^2 \\
          A \lambda^3 (1-\rho-i\eta) & -A \lambda^2 & 1
      \end{array}
    \right),
\end{eqnarray}
the parameter $\eta$ parameterizes the size of the CP violation.

Importantly, the measurements of $\epsilon$ and $\phi_1$ are used to
constrain the parameters in the KM matrix.  Currently, all the
measurements of the CP violations more or less suggest the same region
on the $\rho$ vs.\ $\eta$ plane, and hence the observed CP violations
are well explained in the framework proposed by Kobayashi and Maskawa.
In the near future, some of the measurements of the CP
violations will become more precise \cite{LHCB,KOPIO}.  In addition,
there will be other constraints on the $\rho$ vs.\ $\eta$ plane from
various new processes like $K_L\rightarrow\pi^0\nu\bar{\nu}$ and decay
processes of $B_d$ and $B_s$ mesons.  These processes will provide
important tests of the unitariry of the KM matrix and the origin of
the CP violations.  In particular, if some new physics exists, it may
provide a new source of the CP violation which may be seen as a
deviation from the SM prediction on the $\rho$ vs.\ $\eta$ plane.

Of course, possible deviations depend on featrures of new
physics beyond the standard model.  Among various models, in this
paper, we consider one of the most well-motivated ones, that is,
supersymmetric (SUSY) unified model with right-handed neutrinos.  Such
a model can solve some of the theoretical and experimental problems
which cannot be solved in the framework of SM.  First, in the
supersymmetric theories, the serious naturalness problem can be
avoided because of the cancellation of the quadratic divergences
between bosonic and fermionic loops.  In addition, in the minimal SUSY
SM (MSSM), successful gauge coupling unification can be realized
contrary to the standard-model case where three gauge couplings do not
meet at any high energy scale.  Furthermore, in this framework, solar
and atmospheric neutrino problems \cite{SUPERK,SOLAR} may be solved by
small neutrino masses generated by the seesaw mechanism \cite{seesaw}.
Thus, supersymmetric unified theories with right-handed neutrinos are
theoretically and experimentally well-motivated, and it is worth
studying its phenomenological consequences. In this context there have
been some phenomenological studies \cite{moroi00,baek00} which
obatained interesting results in some processes.

In this paper, we study the CP violating processes in the kaon system
in the framework of supersymmetrics grand unified theories (GUTs) with
right-handed neutrinos.\footnote{Flavor violation in $K$ and $B$
  sysytems in the framework of SUSY SM without the right-handed
  neutrino have been discussed in
  \cite{hall86,kurimoto89,bertolini91,gotookada}. } Indeed, in such
models, there are many possible new physical phases in the soft SUSY
breaking parameters, which may affect various CP violating processes.

This paper is orgenized as follows.  In Section \ref{sec:model}, we
introduce the model we consider.  Then, in Section
\ref{sec:numerical}, rates of the various CP violations in the
$K$-system is numerically evaluated.  Section \ref{sec:conclusion} is
devoted for the conslusions and discussion. Relevant formulae are
collected in the Appendices.

\section{Model}
\label{sec:model}
\setcounter{equation}{0}

In this paper we consider the minimal SU(5) GUT with singlet
right-handed neutrinos.  Let us denote {\bf 10}, $\bar{\textrm{{\bf
      5}}}$, and the right-handed neutino chiral multiplets in $i$-th
generation as $\Psi_i$, $\Phi_i$ and $N_i$, respectively.  The
superpotential is given as follows;\footnote{In addition, an adjoint
  Higgs $\Sigma$ which is responsible for SU(5) breaking is
  assumed. However interactions between $\Sigma$ and the other chiral
  multiplets are not significant for our discussion. Hence we do not discuss
  these interactions in this paper.  }
\begin{eqnarray}
  W_{\rm{GUT}} = \frac{1}{8} \Psi_i [Y_U]_{ij} \Psi_j H
   + \Psi_i [Y_D]_{ij} \Phi_j \bar{H}  +N_i [Y_N]_{ij} \Phi_j H
   + \frac{1}{2} N_i [M_N]_{ij} N_j ,
\end{eqnarray}
where $H$ and $\bar{H}$ are $\rm{{\bf 5}}$ and $\bar{\rm{{\bf
      5}}}$ representation Higgs fields, respectively.  Here $i,j$ are
generation indices.  $M_N$ is Majorana mass matrix of the right-handed
neutrinos.  Throughout the paper we adopt the universal structure of
$M_N$ for simplicity,
\begin{eqnarray}
  [M_N]_{ij} = M_{\nu_R} \delta_{ij} .
  \label{MajoranaMass}
\end{eqnarray}

For our discussion it is convenient to choose basis of $\Psi_i$,
$\Phi_i$ and $N_i$ such that the down-type Yukawa matrix is diagonal
and mixing matrices appear in the up-type and neutrino Yukawa couplings.
The mixing matrix of the up-type Yukawa coupling corresponds to the KM
matrix $V_{\rm KM}$.  In general the mixing matrix of the neutrino Yukawa
coupling is given by a combination of two unitary matrices that are
mixing matrices of left-handed neutrinos and right-handed Majorana
neutrinos.  In our case the mixing matrix of the neutrino Yukawa coupling
becomes the Maki-Nakagawa-Sakata matrix $V_{\rm MNS}$ \cite{maki62}
because of the universal structure given in Eq.\ (\ref{MajoranaMass}).
Hence in this basis Yukawa matrices are decomposed as
\begin{eqnarray}
  Y_U( M_{{\rm GUT}} ) = V_{{\rm KM}}^T \hat \Theta_Q \hat Y_U 
   V_{{\rm KM}}, \quad
  Y_D( M_{{\rm GUT}} ) = \hat Y_D, \quad
  Y_N( M_{{\rm GUT}} ) = \hat Y_N V_{{\rm MNS}}^T \hat \Theta_L,
\end{eqnarray}
where $\hat Y$'s are diagonal matrices;
\begin{eqnarray}
  \hat Y_U = {\rm diag}({y_u}_1, {y_u}_2, {y_u}_3), \quad
  \hat Y_D = {\rm diag}({y_d}_1, {y_d}_2, {y_d}_3), \quad
  \hat Y_N = {\rm diag}({y_n}_1, {y_n}_2, {y_n}_3) ,
\end{eqnarray}
and $\hat \Theta$'s are diagonal phase matrices;
\begin{eqnarray}
  \hat \Theta_Q = {\rm diag}( e^{i \varphi^{(Q)}_1}, e^{i \varphi^{(Q)}_2}, 
   e^{i \varphi^{(Q)}_3} ), \quad
  \hat \Theta_L = {\rm diag}( e^{i \varphi^{(L)}_1}, e^{i \varphi^{(L)}_2}, 
   e^{i \varphi^{(L)}_3} ) .
\end{eqnarray}
Since physics does not change under redefinition of overall phases of
$\hat \Theta_Q$ and $\hat \Theta_L$, we fix $\varphi^{(Q)}_1
+\varphi^{(Q)}_2 +\varphi^{(Q)}_3 = 0$ and $\varphi^{(L)}_1
+\varphi^{(L)}_2 +\varphi^{(L)}_3 = 0$.
$V_{\rm KM}$ and $V_{\rm MNS}$ are parameterized by four parameters, i.e., 
three mixing angles and one phase as follows \cite{PDG};
\begin{eqnarray}
  V_{\rm KM}  = \left(
        \begin{array}{@{\,}cccc@{\,}}
          c^{(Q)}_{12} c^{(Q)}_{13} & s^{(Q)}_{12} c^{(Q)}_{13} 
           & s^{(Q)}_{13} e^{-i \delta^{(Q)}_{13}} \\
          -s^{(Q)}_{12} c^{(Q)}_{23} 
            -c^{(Q)}_{12} s^{(Q)}_{23} s^{(Q)}_{13} e^{i \delta^{(Q)}_{13}}
           & c^{(Q)}_{12} c^{(Q)}_{23} 
            -s^{(Q)}_{12} s^{(Q)}_{23} s^{(Q)}_{13} e^{i \delta^{(Q)}_{13}}
           & s^{(Q)}_{23} c^{(Q)}_{13} \\
          s^{(Q)}_{12} s^{(Q)}_{23}
            -c^{(Q)}_{12} c^{(Q)}_{23} s^{(Q)}_{13} e^{i \delta^{(Q)}_{13}}
           & -c^{(Q)}_{12} s^{(Q)}_{23} 
            -s^{(Q)}_{12} c^{(Q)}_{23} s^{(Q)}_{13} e^{i \delta^{(Q)}_{13}}
           & c^{(Q)}_{23} c^{(Q)}_{13}
        \end{array}
                 \right) ,\quad
\end{eqnarray}
where $c^{(Q)}_{ij} = \cos \theta^{(Q)}_{ij}$ and $s^{(Q)}_{ij} = \sin
\theta^{(Q)}_{ij}$, and $V_{\rm MNS}$ is obtained by exchanging $Q
\leftrightarrow L$.

Let us decompose $W_{{\rm GUT}}$ by using the standard model fields.
We embed the standard model fields into the GUT fields as
\begin{eqnarray}
\Psi_i^{AB} = \left(
        \begin{array}{@{\,}cccc@{\,}}
          V_{{\rm KM}}^\dagger \hat \Theta_Q^\dagger 
           \epsilon^{\alpha \beta \gamma} \bar{U}_\gamma 
         
          & -Q^{\alpha a} \\ 
          Q^{\alpha a}
          & - \hat\Theta_L \epsilon^{ab} \bar{E} \\
        \end{array}
                  \right) ,\quad
  \Phi_{iA} = \left(
        \begin{array}{@{\,}cccc@{\,}}
          \bar{D}_\alpha \\
          \hat \Theta_L^\dagger \epsilon_{ab} L^b
        \end{array}
                     \right), 
  \label{eq:embeding}
\end{eqnarray}
where $Q_i ({\bf 3}, {\bf 2})_{1/6}$, $\bar{U}_i ({\bf \bar{3}}, {\bf
1})_{-2/3}$, $\bar{D}_i ({\bf \bar{3}}, {\bf 1})_{1/3}$, $L_i ({\bf 1}, {\bf
2})_{-1/2}$ and $\bar{E}_i ({\bf 1}, {\bf 1})_{1}$ are quark and lepton
fields in $i$-th generation with ${\rm SU(3)}_C \times {\rm SU(2)}_L
\times {\rm U(1)}_Y$ quantum numbers as indicated.  Here
$\alpha,\beta,\gamma$ represent ${\rm SU(3)}_C$ indices and $a,b$ are
${\rm SU(2)}_L$ indices.  With this embedding, $W_{{\rm GUT}}$ at the
GUT scale becomes
\begin{eqnarray}
  W_{{\rm GUT}} &=& W_{{\rm SSM}} 
   -\frac{1}{2}Q_i [V_{{\rm KM}}^T \hat \Theta_Q \hat Y_U 
    V_{{\rm KM}}]_{ij} Q_j H_C
   -\bar{E}_i [\hat \Theta_L \hat Y_U]_{ij} \bar{U}_j H_C
   +\bar{U}_i [\hat \Theta_Q^\dagger V_{{\rm KM}}^* \hat Y_D]_{ij} 
    \bar{D}_j \bar{H}_C \nonumber \\
   & &
   -Q_i [\hat Y_D \hat \Theta_L]_{ij} L_j \bar{H}_C
   +N_i [\hat Y_N V_{{\rm MNS}}^T \hat \Theta_L]_{ij} \bar{D}_j H_C,
 \label{eq:WGUT}
\end{eqnarray}
where $H_C$ and $\bar{H}_C$ are colored Higgs fields, and low energy
superpotential is given as
\begin{eqnarray}
  W_{{\rm SSM}} &=& H_u \bar{U}_i [\hat{Y}_U V_{\rm KM}]_{ij} Q_j
   + H_d \bar{D}_i [\hat Y_D]_{ij} Q_j
   + H_d \bar{E}_i [\hat Y_E]_{ij} L_j \nonumber \\
   & &
   +N_i [\hat Y_N V_{{\rm MNS}}^T ]_{ij} L_j H_u
   +\frac{1}{2} M_{\nu_R} N_i N_i .
   \label{eq:WSSM}
\end{eqnarray}
Notice that the phase matrices $\hat \Theta_{Q,L}$ do not appear in
Eq.(\ref{eq:WSSM}).  Hence these phases $\varphi^{(Q,L)}_i$ are
independent of the CP phase in the KM matrix.  As we will show later,
GUT phases $\varphi^{(L)}_i$ can have important implications to CP
violations in the kaon system.  Although simple SU(5) GUT predicts $\hat{Y}_D =
\hat{Y}_E$ at the GUT scale, this relation is unrealistic for the
first and second generations.  In order to explain realistic fermion
mass pattern, some new flavor physics are necessary.  Although these
new physics can provide extra source of flavor and CP violations 
\cite{arkanihamed96,ciafaloni96,HNOST98}, we do not consider these
effects in this paper.

With the superpotential given in Eq.(\ref{eq:WSSM}), the left-handed neutrino 
mass matrix is generated by the seesaw mechanism,
\begin{eqnarray}
  [m_{\nu_L}]_{ij} = \frac{ \langle H_u \rangle^2}{M_{\nu_R}} 
   [V_{{\rm MNS}} \hat Y_N^2 V_{{\rm MNS}}^T]_{ij}
   = \frac{v^2 \sin^2 \beta}{2 M_{\nu_R}} 
   [V_{{\rm MNS}} \hat Y_N^2 V_{{\rm MNS}}^T]_{ij} , 
   \label{eq:nuLmass}
\end{eqnarray}
where $v\simeq246$ GeV and $\tan \beta$ is the ratio of two Higgs
vacuum expectation values.  In our analysis, we use the masses and
mixing angles of neutrinos suggested by solar and atmospheric neutrino
data.  Atmospheric neutrino data \cite{SUPERK} at Super-Kamiokande
implies that $\nu_\mu - \nu_\tau$ mixing is almost maximal, hence we
set $s^{(L)}_{23} \simeq 1/\sqrt{2}$.  As for $\nu_{e} - \nu_{\mu}$
mixing, we consider large and small angle MSW solutions to the solar
neutrino problem\cite{BilenkyGiunti}.\footnote{ There are other
  solutions to the solar neutrino problem, LOW, vacuum and Just-So
  solutions. In these solutions, second generation neutrino mass is
  smaller than that in large or small angle MSW solutions.  $m_{\nu_2}
  \sim 3 \times 10^{-4} {\rm eV}$, $1.2 \times 10^{-5} {\rm eV}$ and
  $2 \times 10^{-6} {\rm eV}$ for LOW, vacuum and Just-So solutions,
  respectively.  When we impose perturbativity of $y_{n_3}$, $y_{n_2}$
  is very small because of the universality of $M_N$.  Therefore when
  $s^{(L)}_{13} = 0$, flavor violation for the kaon system is too
  small to be observed.  So we do not discuss the three cases.  But if
  $M_N$ is not universal or $s^{(L)}_{13}$ is nonzero, the situation
  may change.} For large angle MSW solution, we use
\begin{eqnarray}
  m_{\nu} \simeq (0, ~0.004{\rm eV}, ~0.03{\rm eV}), \quad
  V_{{\rm MNS}} \simeq \left(
        \begin{array}{@{\,}cccc@{\,}}
           0.91 & -0.42 & \ll 1 \\
          -0.30 &  0.64 & 0.71 \\
           0.30 & -0.64 & 0.70
        \end{array}
        \right) ,
\end{eqnarray}
and for small angle MSW solution, 
\begin{eqnarray}
  m_{\nu} \simeq (0, ~0.003{\rm eV}, ~0.03{\rm eV}), \quad
  V_{{\rm MNS}}\simeq \left(
        \begin{array}{@{\,}cccc@{\,}}
          1.0   & 0.040 & \ll 1 \\
         -0.028 &  0.70 & 0.71 \\
          0.028 & -0.71 & 0.70
        \end{array}
        \right) .
\end{eqnarray}
$[V_{\rm MNS}]_{13}$ is small as indicated by CHOOZ experiment
\cite{CHOOZ}.  In our analysis we set $[V_{\rm MNS}]_{13} = 0$ unless
otherwise stated.  Nonzero value of $[V_{\rm MNS}]_{13}$ can
affect our result, especially for the small angle MSW case as we will
discuss later.

In our basis, Yukawa matrices for down-type quarks and charged leptons
are diagonal at the GUT scale.  Their diagonalities
do not hold at the weak scale because of renormalization group (RG)
effects.  However these effects are minor for our discussion and
$\bar{D}_i$ and $L_i$ fields in Eq.(\ref{eq:WSSM}) are mass
eigenstates with good accuracy.  Hence it is very useful to use this
basis to see the flavor violations in $K$ and $B$ meson system caused
by the right-handed neutrinos.

Although mass matrices of down-type quarks and charged leptons are
diagonal in our basis, down-type squark mass matrix may have flavor
violating off-diagonal elements.  The soft SUSY breaking terms for
scalar fields are given by
\begin{eqnarray}
  V_{\rm GUT} &=& 
     \frac{1}{2} \tilde{\psi}_i [m_{10}^2]_{ij} \tilde{\psi}^\dagger_j
     +\tilde{\phi}^\dagger_i [m_5^2]_{ij} \tilde{\phi}_j
     +\tilde{n}_i [m_N^2]_{ij} \tilde{n}^\dagger_j
     +m_H^2 h^\dagger h
     +m_{\bar{H}}^2 \bar{h}^\dagger \bar{h} \nonumber \\
    & &
     +\frac{1}{8} \tilde{\psi}_i [A_U]_{ij} \tilde{\psi}_j h 
     +\tilde{\psi}_i [A_D]_{ij} \tilde{\phi}_j \bar{h}
     +\tilde{n}_i [A_N]_{ij} \tilde{\phi}_j h,
\end{eqnarray}
where $\tilde{\psi}_i$, $\tilde{\phi}_i$, $\tilde{n}_i$, $h$ and
$\bar{h}$ are scalar components of the superfields $\Psi_i$, $\Phi_i$,
$N_i$, $H$ and $\bar{H}$, respectively.  Soft SUSY breaking terms for
the low energy effective theory below the GUT scale are given by
\begin{eqnarray}
  V_{\rm SSM} &=& \tilde{q}^\dagger_i [m_Q^2]_{ij} \tilde{q}_j
     +\tilde{\bar{d}}_i [m_D^2]_{ij} \tilde{\bar{d}}_j^\dagger
     +\tilde{\bar{u}}_i [m_U^2]_{ij} \tilde{\bar{u}}_j^\dagger
     +\tilde{l}^\dagger [m_L^2]_{ij} \tilde{l}_j
     +\tilde{\bar{e}}_i [m_E^2]_{ij} \tilde{\bar{e}}_j^\dagger 
     +\tilde{n} [m_N^2] \tilde{n}_j^\dagger \nonumber \\
    & & +h_u \tilde{\bar{u}}_i [A_u]_{ij} \tilde{q}_j
     + h_d \tilde{\bar{d}}_i [A_d]_{ij} \tilde{q}_j
     + h_d \tilde{\bar{e}}_i [A_e]_{ij} \tilde{l}_j
     + h_u \tilde{n}_i [A_n]_{ij} \tilde{l}_j,
\end{eqnarray}
where $\tilde{q}_i$, $\tilde{\bar{u}}_i$, $\tilde{\bar{d}}_i$,
$\tilde{l}_i$, $\tilde{\bar{e}}_i$, $\tilde{n}_i$ $h_u$ and $h_d$ are
scalar components of the superfields $Q_i$, $\bar{U}_i$, $\bar{D}_i$,
$L_i$, $\bar{E}_i$, $N_i$, $H_u$ and $H_d$, respectively.  With the embedding
Eq.(\ref{eq:embeding}) the above scalar mass matrices at the GUT scale
are given by
\begin{eqnarray}
  & &
  m_Q^2 = {m_{10}^2}^T, \quad
  m_U^2 = \hat \Theta_Q^\dagger V_{{\rm KM}}^* m_{10}^2 V_{{\rm KM}}^T
   \hat \Theta_Q , \quad
  m_D^2 = {m_5^2}^T, \nonumber \\
  & &
  m_L^2 = \hat{\Theta}_L m_5^2 {\hat{\Theta}_L}^\dagger, \quad
  m_E^2 = \hat{\Theta}_L m_{10}^2 {\hat{\Theta}_L}^\dagger .
\end{eqnarray}

We are particularly interested in the
RG effect on the off-diagonal elements of the sfermion mass matrices.
Thus, in our analysis, we use the boundary condition of the minimal
supergravity model.  Then, the SUSY breaking parameters at the reduced
Planck scale $M_* \simeq 2.4 \times 10^{18} {\rm GeV}$ are give by
\begin{eqnarray}
  & & m_{10}^2(M_*) = m_5^2(M_*) 
  = m_N^2(M_*) = m_0^2 {\bf 1}, \nonumber \\
  & & A_U(M_*) = a_0 Y_U(M_*),\quad  
  A_D(M_*) = a_0 Y_D(M_*), \quad  A_N(M_*) = a_0 Y_N(M_*), 
   \nonumber \\
  & & M_1(M_*) = M_2(M_*) = M_3(M_*) = M_{1/2},
\end{eqnarray}
where $M_1$, $M_2$ and $M_3$ are SUSY breaking masses of U(1)$_Y$,
SU(2)$_L$ and SU(3)$_C$ gauginos, respectively.  Importantly, sfermion
masses are universal at the reduced Planck scale.  Below the Planck
scale $M_*$, however, off-diagonal elements of scalar mass matrices
are generated by the RG effect and sizable CP and flavor violations
may be possible.  Of course, if there are tree-level off-diagonal
elements of the sfermion mass matrices at the scale $M_*$, they may
affect the rates of the CP and flavor violating processes.  However,
it is fairly unnatural to assume cancellations between the tree-level
and RG contributions to the off-diagonal elements.  Thus, we
believe our approach will give us a conservative estimation of the rate
of the CP and flavor violating processes unless some accidental
cancellation happens.

First we see lepton flavor violation.  In the presence of the
right-handed neutrinos, $H_u - N - L$ interaction, the fourth term in
Eq.(\ref{eq:WSSM}), violates lepton flavor.  Although the right-handed
neutrinos are very heavy, this interaction affects the RG evolution of
slepton mass matrix and generates off-diagonal elements of $m_L^2$ at
the weak scale.  Such an effect can be probed by $\mu \to e \gamma$,
$\tau \to \mu \gamma$, $\mu-e$ conversion, and so on
\cite{borzumati86,HMTYNR,HisanoNomura}.\footnote{In the GUT framework,
  lepton flavor is violated by interaction with colored Higgs. Such a
  effect have been studied in \cite{ciafaloni96,barbierihall,HMTYSU5}}

Since down-type quarks and charged leptons belong to the same
representation of SU(5), flavor of down-type quarks is also violated
via the right-handed neutrino Yukawa interaction
\cite{BarbieriHallStrumia,moroi00,baek00}.  Indeed interaction among
$N$, $\bar{D}$ and $H_C$ (the last term of Eq.(\ref{eq:WGUT})) induces
off-diagonal elements of the right-handed down-type squark mass
matrix, which are evaluated with the one iteration approximation as
\begin{eqnarray}
  [m_D^2]_{ij} \simeq \frac{1}{8 \pi^2} (3m_0^2 +a_0^2) 
   e^{i(\varphi^{(L)}_i-\varphi^{(L)}_j)} 
   \sum_k {y_{n}}_k^2 [V_{{\rm MNS}}]_{ik} [V_{{\rm MNS}}^*]_{jk}
   \log \frac{M_{{\rm GUT}}}{M_*} \quad (i \neq j).
\end{eqnarray}
Notice that the size of the off-diagonal elements is determined by the
parameters in the neutrino sector, i.e., neutrino Yukawa coupling and
$V_{{\rm MNS}}$.  In addition these elements depend on phase
$\varphi^{(L)}_i-\varphi^{(L)}_j$.  Hence flavor violation and new
phases caused by the right-handed neutrino affect flavor and CP
violations in the quark sector.  In this paper we focus on kaon system
to investigate such effects.

In order to discuss flavor violation, it is useful to introduce the
following variable which is off-diagonal elements of $m_D^2$
normalized by squark mass scale $m_{\tilde{q}}$,
\begin{eqnarray}
  \Delta^{(R)}_{ij} \equiv \frac{ [m_D^2]_{ij} }{ m_{\tilde{q}}^2 }
   = \frac{ [m_D^2]_{ij} }{ [m_D^2]_{11} }.
\end{eqnarray}
Here we choose the first generation down-type squark mass as squark
mass scale, which is approximately estimated as $[m_D^2]_{11} \simeq
m_0^2 +7.1M_{1/2}^2$ in our model.  Using $s^{(L)}_{23} \simeq
1/\sqrt{2}$ and $ (s^{(L)}_{13})^2 \ll 1$, $\Delta^{(R)}_{12}$ is
approximately estimated as
\begin{eqnarray}
  \Delta^{(R)}_{12} &\simeq& 5.5 \times 10^{-4}
\nonumber \\ && \times
  \left( \frac{M_{\nu_R}}{10^{14} {\rm GeV}} \right)
  \left( \frac{m_{\nu_2}}{0.004 {\rm eV}} \right)
  \frac{ e^{i (\varphi^{(L)}_1 - \varphi^{(L)}_2) } }{\sin^2 \beta}
  \frac{ 3m_0^2 +a_0^2 }{m_{\tilde{q}}^2}
  \left( s^{(L)}_{12} c^{(L)}_{12} 
    +\frac{ m_{\nu_3} }{ m_{\nu_2} } s^{(L)}_{13} \right).
  \label{eq:approx_DelR12}
\end{eqnarray}
From the equation, we find that the flavor violation is enhanced for
larger $M_{\nu_R}$.  It reflects one characteristic feature of the seesaw
mechanism, $y_{n_i} \propto \sqrt{M_{\nu_R}}$ with fixed left-handed
neutrino mass.

$\Delta^{(R)}_{12}$ is sensitive to the neutrino mixing angles.  First
we discuss $s^{(L)}_{13} = 0$ case.  For the large angle MSW solution,
$\Delta^{(R)}_{12}$ is not suppressed by mixing angle because
$s^{(L)}_{12} c^{(L)}_{12} \sim 1/2$.  On the other hand,
$\Delta^{(R)}_{12}$ is suppressed by the factor $s^{(L)}_{12}
c^{(L)}_{12} \sim 0.04$ for the small angle MSW case.  Hence, for fixed
$M_{\nu_R}$, amplitudes for flavor violating processes in the large angle
MSW case are larger than those in the small angle case by one order of
magnitude.  Next we consider an effect of finite $s^{(L)}_{13}$.
Since $m_{\nu_3} / m_{\nu_2} \sim 10$, the effect becomes important
when $ s^{(L)}_{13} \gtrsim 0.04$ for the large angle MSW case, and $
s^{(L)}_{13} \gtrsim 0.004$ for the small angle MSW case.  Therefore
improvement of the upper bound on $s^{(L)}_{13}$ can have impact on
the quark sector flavor and CP violations, especially for the small angle
MSW case, in our model.

Off-diagonal elements of the left-handed squark mass matrix
are generated by quark flavor mixing via $V_{\rm KM}$,
\begin{eqnarray}
  [m_Q^2]_{ij} = \frac{1}{8 \pi^2} (3m_0^2 +a_0^2) 
   y_t^2 [V_{{\rm KM}}]_{3i}^* [V_{{\rm KM}}]_{3j}
   \left( 3\log \frac{M_{{\rm GUT}}}{M_*} 
     +\log \frac{{M_{{\rm weak}}}}{M_{{\rm GUT}}} \right) 
   \quad (i \neq j).
\end{eqnarray}
As in the $\tilde{d}_R$ sector, we introduce the following variable
\begin{eqnarray}
  \Delta^{(L)}_{ij} \equiv \frac{ [m_Q^2]_{ij} }{ m_{\tilde{q}}^2 }.
\end{eqnarray}
$\Delta^{(L)}_{12}$ is approximately given by
\begin{eqnarray}
\Delta^{(L)}_{12} \simeq 1.6 \times 10^{-4} \frac{1}{\sin^2 \beta} 
\frac{3 m_0^2 +a_0^2}{m_{\tilde{q}}} e^{i \phi_1} .
\end{eqnarray}
Comparing the above expression to Eq.(\ref{eq:approx_DelR12}), we find that
$\Delta^{(R)}_{12}$ dominates over $\Delta^{(L)}_{12}$ when $M_{\nu_R}
\gtrsim 10^{14}$ GeV.

\section{Numerical Results}
\label{sec:numerical}
\setcounter{equation}{0}

\subsection{$\epsilon$ and $\Delta m_K$}

In this subsection we discuss effects of the right-handed neutrinos on
the $\Delta S =2$ processes.

The effective Hamiltonian for the $\Delta S = 2$ processes is given by
\begin{eqnarray}
  {\mathcal H}_{\rm eff} = \sum_{i=1}^3 [ C_{L,i}(\mu) {\mathcal Q}_{L,i}(\mu) 
   + C_{R,i}(\mu) {\mathcal Q}_{R,i}(\mu) ]
  + \sum_{i=4}^5 C_i(\mu) {\mathcal Q}_i(\mu) ,
  \label{eq:Heff_DelS2}
\end{eqnarray}
where operators are 
\begin{eqnarray}
  {\mathcal Q}_{L,1} &=& 4( \bar{d}_\alpha \gamma_\mu P_L s_\alpha )
   ( \bar{d}_\beta \gamma^\mu P_L s_\beta) ,  \nonumber \\
  {\mathcal Q}_{L,2} &=& 4 ( \bar{d}_\alpha P_L s_\alpha )
   ( \bar{d}_\beta P_L s_\beta ) , \nonumber \\
  {\mathcal Q}_{L,3} &=& 4 ( \bar{d}_\alpha P_L s_\beta )  
   ( \bar{d}_\beta P_L s_\alpha ), \nonumber \\
  {\mathcal Q}_4 &=& 4 ( \bar{d}_\alpha P_L s_\alpha) 
   ( \bar{d}_\beta P_R s_\beta ), \nonumber \\
  {\mathcal Q}_5 &=& 4 ( \bar{d}_\alpha P_L s_\beta )  
   ( \bar{d}_\beta P_R s_\alpha ),
  \label{eq:DelS2OP}
\end{eqnarray}
and the operators ${\mathcal Q}_{R,i}$ ($i=1-3$) are obtained from
${\mathcal Q}_{L,i}$ by exchanging $ L \leftrightarrow R $.  Here
$\alpha$ and $\beta$ are color indices.  In order to calculate $\Delta
m_K$ and $\epsilon$, we take account of QCD correction to the Wilson
coefficients, and using formulae for the
leading-order QCD correction given in \cite{Matchev97}, the matrix
element of the effective Hamiltonian at $\mu_c =1.3$ GeV is given as
\begin{eqnarray}
  \langle K^0 | {\mathcal H}_{\rm eff} | \bar{K}^0 \rangle &=&
  \eta_1 \left[ C_{1,L} 
               +C_{1,R} \right] 
   \frac{2}{3} m_K^2 f_K^2 \nonumber \\
  & & +\left[ \eta_4 C_4 
             +\frac{1}{3}(\eta_4 -\eta_5) C_5 \right]
   \left[ \frac{1}{12} +\frac{1}{2}\left(\frac{m_K}{m_s +m_d}\right)^2 
\right]
   m_K^2 f_K^2  \nonumber \\
  & & +\eta_5 C_5 \left[ \frac{1}{4} 
    +\frac{1}{6} \left(\frac{m_K}{m_s+m_d} \right)^2 \right] m_K^2 f_K^2 ,
  \label{eq:Heff}
\end{eqnarray}
where we used vacuum saturation approximation to calculate matrix
elements, and the QCD correction factors are:
\begin{eqnarray}
  \eta_1 = \left( \frac{\alpha_s(m_b)}{\alpha_s(m_c)} \right)^{6/25} 
          \left( \frac{\alpha_s(m_W)}{\alpha_s(m_b)} \right)^{6/23}
         \simeq 0.77,
\end{eqnarray}
and $\eta_4=\eta_1^{-4}\simeq2.8$ and $\eta_5=\eta_1^{1/2}\simeq0.88$.

With this matrix element, $\epsilon$ and $\Delta m_K$ are
given by
\begin{eqnarray}
  \epsilon &=& \frac{ e^{i\pi/4} {\rm Im} 
    \langle K^0 | {\mathcal H}_{\rm eff} | \bar{K}^0 \rangle }
   {2\sqrt{2} m_K \Delta m_K} ,  \\
  \Delta m_K &=& \frac{1}{m_K} 
   |\langle K^0 | {\mathcal H}_{\rm eff} | \bar{K}^0 \rangle |  .
\end{eqnarray}

There are four kinds of the SUSY contributions to the effective
Hamiltonian, 1) gluino mediated, 2) chargino mediated, 3)
gluino-neutralino mediated and 4) neutralino mediated.  Among them,
gluino mediated $O(\alpha_s^2)$ contribution dominates over the others
in our model.  Hence for a while we concentrate on the gluino
contribution for an intuitive discussion.  At the scale where SUSY
particles are integrated out, Wilson coefficients are given in terms
of mass insertion (MI) parameters as follows \cite{MassInsertion}:
\begin{eqnarray}
  \tilde{C}_{L,1} |^{\tilde{G}}_{\rm MI} &=& 
  \frac{4 \pi^2}{9} \alpha_s^2 
  \left( \Delta^{(L)}_{12} \right)^2 m_{\tilde{q}}^4
  [ 4 m_{\tilde{G}}^2 I^0_6(m_{\tilde{G}}^2,m_{\tilde{G}}^2,
  m_{\tilde{q}}^2,m_{\tilde{q}}^2,m_{\tilde{q}}^2,m_{\tilde{q}}^2) 
  \nonumber \\
  & &+11 I^1_6(m_{\tilde{G}}^2,m_{\tilde{G}}^2,
  m_{\tilde{q}}^2,m_{\tilde{q}}^2,m_{\tilde{q}}^2,m_{\tilde{q}}^2) ], \\
  \tilde{C}_{R,1} |^{\tilde{G}}_{\rm MI} &=& \frac{4 \pi^2}{9} \alpha_s^2 
  \left( \Delta^{(R)}_{12} \right)^2 m_{\tilde{q}}^4
  [ 4 m_{\tilde{G}}^2 I^0_6(m_{\tilde{G}}^2,m_{\tilde{G}}^2,
  m_{\tilde{q}}^2,m_{\tilde{q}}^2,m_{\tilde{q}}^2,m_{\tilde{q}}^2) 
  \nonumber \\
  & &+11 I^1_6(m_{\tilde{G}}^2,m_{\tilde{G}}^2,
  m_{\tilde{q}}^2,m_{\tilde{q}}^2,m_{\tilde{q}}^2,m_{\tilde{q}}^2) ], \\
  \tilde{C}_{4} |^{\tilde{G}}_{\rm MI} &=& \frac{16 \pi^2}{3} \alpha_s^2
  \left( \Delta^{(L)}_{12} \right) \left( \Delta^{(R)}_{12} \right) 
  m_{\tilde{q}}^4
  [ 7 m_{\tilde{G}}^2 I^0_6(m_{\tilde{G}}^2,m_{\tilde{G}}^2,
  m_{\tilde{q}}^2,m_{\tilde{q}}^2,m_{\tilde{q}}^2,m_{\tilde{q}}^2) 
  \nonumber \\
  & & -I^1_6(m_{\tilde{G}}^2,m_{\tilde{G}}^2,
  m_{\tilde{q}}^2,m_{\tilde{q}}^2,m_{\tilde{q}}^2,m_{\tilde{q}}^2) ], \\
  \tilde{C}_{5} |^{\tilde{G}}_{\rm MI} &=& \frac{16 \pi^2}{9} \alpha_s^2
  \left( \Delta^{(L)}_{12} \right) \left( \Delta^{(R)}_{12} \right) 
  m_{\tilde{q}}^4
  [ m_{\tilde{G}}^2 I^0_6(m_{\tilde{G}}^2,m_{\tilde{G}}^2,
  m_{\tilde{q}}^2,m_{\tilde{q}}^2,m_{\tilde{q}}^2,m_{\tilde{q}}^2) 
  \nonumber \\
  & & +5 I^1_6(m_{\tilde{G}}^2,m_{\tilde{G}}^2,
  m_{\tilde{q}}^2,m_{\tilde{q}}^2,m_{\tilde{q}}^2,m_{\tilde{q}}^2) ],
\end{eqnarray}
where the function $I^N_D$ is defined in the Appendix \ref{app:I}.
Since flavor violating left-right mixing is very small, so are
$\tilde{C}_{L,2} |^{\tilde{G}}_{\rm MI}$, $\tilde{C}_{R,2}
|^{\tilde{G}}_{\rm MI}$, $\tilde{C}_{L,3} |^{\tilde{G}}_{\rm MI}$,
$\tilde{C}_{R,3} |^{\tilde{G}}_{\rm MI}$.  Therefore we neglect them
in the following rough estimation.  Although these coefficients are
suppressed by $1/m_{\tilde{q}}^2$, they can be comparable to the SM
contributions since they are $O(\alpha_s^2)$ while the SM
contributions are $O(\alpha_2^2)$.  Notice that contributions from
operators ${\mathcal Q}_{R,1}$, ${\mathcal Q}_4$ and ${\mathcal Q}_5$
are sizable only when flavor in $\tilde{d}_R$ sector is violated.

The flavor and CP violations in $\tilde{d}_R$ sector have important
implication to $\Delta m_K$ and $\epsilon$ because the matrix elements
of the $LR$ type operators ${\mathcal Q}_4$ and ${\mathcal Q}_5$ are
enhanced by the factor $(m_K/m_s)^2 \sim 10$.  In addition the Wilson
coefficient of the operator ${\mathcal Q}_4$ at $\mu_c$ is enhanced by
the QCD correction factor $\eta_4$.  Hence contribution from the
operator ${\mathcal Q}_4$ dominates over the other SUSY contributions
when $\Delta^{(L)}_{12}$ and $\Delta^{(R)}_{12}$ are comparable.  This
is the case for the minimal SU(5) GUT with the right-handed neutrinos, and
we shall see that SUSY contribution to $\epsilon$ can be as large as
the experimental value.  In the absence of the right-handed neutrinos,
the $LR$ type operators do not exit, and SUSY contribution to
$\epsilon$ and $\Delta m_K$ is more suppressed.

Before showing numerical results, let us discuss how $\epsilon$ and
$\Delta m_K$ depend on model parameters.  To obtain approximate
formula, we consider only the contribution from ${\mathcal Q}_4$ since
this operator has the most important effect.  In addition, for
simplicity, we consider the case where the gluino is as heavy as
squarks,
Then SUSY contributions  to $\epsilon$ and $\Delta m_K$
are estimated as
\begin{eqnarray}
  \epsilon^{\rm SUSY} &\simeq& 3 \times 10^{-3} e^{i \pi/4}
   \left( \frac{1 {\rm TeV}}{m_{\tilde{q}}} \right)^2
   \left( \frac{ M_{\nu_R} }{ 10^{14} {\rm GeV} } \right)
   \left( \frac{ m_{\nu_2} }{ 0.004 {\rm eV} } \right)
   \left( s^{(L)}_{12} c^{(L)}_{12} +\frac{m_{\nu_3}}{m_{\nu_2}} s^{(L)}_{13}
    \right) 
    \nonumber \\
   & & \times
   \frac{ \sin \left( \varphi^{(L)}_1 - \varphi^{(L)}_2 +\phi_1 \right) }
    {\sin^4 \beta } ,
   \label{eq:rough_epsilon} \\
  \Delta m_K^{\rm SUSY} &\simeq& 2 \times 10^{-14} {\rm MeV}
  \left( \frac{1 {\rm TeV}}{m_{\tilde{q}}} \right)^2
  \left( \frac{ M_{\nu_R} }{ 10^{14} {\rm GeV} } \right)
  \left( \frac{ m_{\nu_2} }{ 0.004 {\rm eV} } \right)
   \nonumber \\
  & & \times 
  \frac{1}{\sin^4 \beta}
  \left( s^{(L)}_{12} c^{(L)}_{12} +\frac{m_{\nu_3}}{m_{\nu_2}} 
    s^{(L)}_{13} \right) .
\end{eqnarray}
From the above estimation, we find that $\epsilon^{\rm SUSY}$ can be
as large as experimental value $\epsilon^{\rm exp} = (2.271 \pm 0.017)
\times 10^{-3}$ \cite{PDG} if $M_{\nu_R} \gtrsim 10^{14} {\rm GeV}$,
for $m_{\tilde{q}} \simeq 1 {\rm TeV}$.  On the other hand $\Delta
m_K^{{\rm SUSY}}$ is much smaller than the experimental value $\Delta
m_K^{{\rm exp}} = (3.489 \pm 0.008) \times 10^{-12} {\rm MeV}$
\cite{PDG} unless $M_{\nu_R}$ is larger than about $10^{15}$ GeV.  We
do not consider such a large right-handed neutrino scale since the
neutrino Yukawa coupling $[Y_N]_{33}$ blows up below the Planck scale
for $M_{\nu_R} \gtrsim 2 \times 10^{15} {\rm GeV}$.  Hence the CP
violation in $\tilde{d}_R$ sector can affect $\epsilon$ and does not
contradict to $\Delta m_K$ measurement.

Now we show our numerical results.  In the numerical calculation not
only the gluino contribution, but also other contributions with chargino and
neutralino propagators are included.  Furthermore contributions from all eight
operators shown in Eq.(\ref{eq:Heff_DelS2}) are taken into account.
Numerically we checked that the gluino contribution through the
operator ${\mathcal Q}_4$ is dominant SUSY contribution.

\begin{figure}
\begin{center}
\leavevmode 
\psfig{file=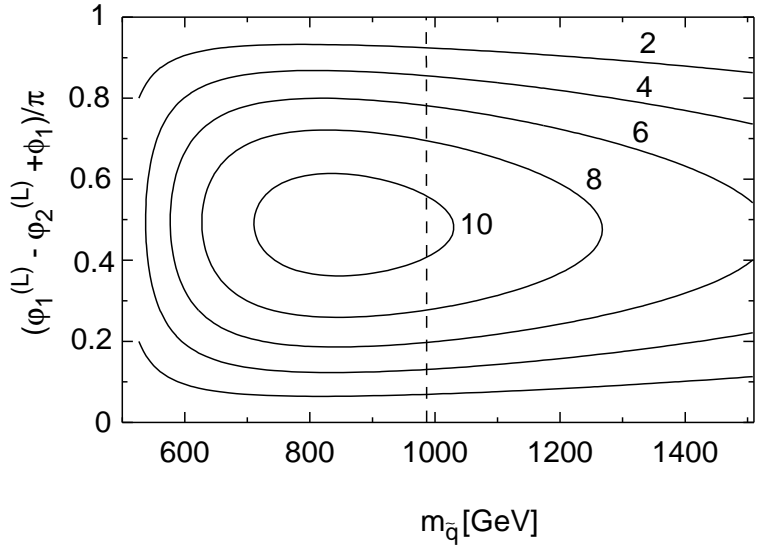,width=10cm}
\end{center}
\caption{Contour of $|\epsilon^{\rm SUSY}|$ (in units of $10^{-3}$) on
$m_{\tilde{q}}$ vs. $(\varphi^{(L)}_1 - \varphi^{(L)}_2 - \phi_1)/\pi$
plane for the large angle MSW case (solid lines).  We take
$M_2 = 150 {\rm GeV}$, $M_{\nu_R} = 3 \times 10^{14} {\rm
GeV}$, $a_0 = 0$ and $\tan \beta = 3$.  In the left side of the dashed
line, $Br(\mu \to e \gamma)$ is larger than experimental limit $1.2
\times 10^{-11}$, provided that $\hat{Y}_D = \hat{Y}_E$ at the GUT
scale.}
\label{fig1}
\end{figure}

In Fig. \ref{fig1}, we show $\epsilon^{{\rm SUSY}}$ on $m_{\tilde{q}}$
vs.  $ \varphi^{(L)}_1 - \varphi^{(L)}_2 + \phi_1 $ plane for large
angle MSW solution with $M_{\nu_R} = 3 \times 10^{14} {\rm GeV}$,
$\tan \beta = 3$, $M_2=150 {\rm GeV}$ and $a_0 = 0$.  From the figure,
we find that $\epsilon^{{\rm SUSY}}$ is maximized at $ \varphi^{(L)}_1
- \varphi^{(L)}_2 + \phi_1 \simeq \pi /2$, as suggested by the
approximation Eq.(\ref{eq:rough_epsilon}).  Notice that heavy Majorana
neutrino mass and large mixing angle induce large lepton flavor
violation in $e$ and $\mu$ sector as well in our model.  Since such
lepton flavor violation is most severely constrained by $\mu \to e
\gamma$, we also calculate $Br(\mu \to e \gamma)$ assuming $\hat{Y}_D
= \hat{Y}_E$ at the GUT scale as the simple SU(5) model predicts.  For
the above parameters, $Br(\mu \to e \gamma)$ is larger than the
experimental upper limit $1.2 \times 10^{-11}$ unless $m_{\tilde{q}}
\gtrsim 980 {\rm GeV}$ as shown in Fig.\ \ref{fig1}.  $\epsilon^{\rm
  SUSY}$ does not depend on $\tan \beta$ so much.  On the other hand,
the constraint gets severer for larger $\tan \beta$ since the
amplitude for $\mu \to e \gamma$ is almost proportional to $\tan
\beta$.  Notice that, even with relatively heavy squark mass of
$m_{\tilde{q}}\gtrsim 980\ {\rm GeV}$ which is consistent with the
$\mu\rightarrow e\gamma$ constraint, the SUSY contribution may be as
large as (or even larger than) the experimentally measured value.  In
addition, the $\mu\rightarrow e\gamma$ constraint here is based on the
relation $\hat{Y}_D = \hat{Y}_E$ at the GUT scale, this relation may
be violated by some new flavor physics which gives realistic fermion
mass pattern for the first and second generations.  Such a mechanism
can change the prediction for $\mu \to e \gamma$ branching ratio from
that in the simple SU(5) case
\cite{ciafaloni96,arkanihamed96,HNOST98}.  Hence we should keep in
mind that this constraint have uncertainty stemming from nontrivial texture
of the fermion mass matrices.

\begin{figure}
\begin{center}
\leavevmode 
\psfig{file=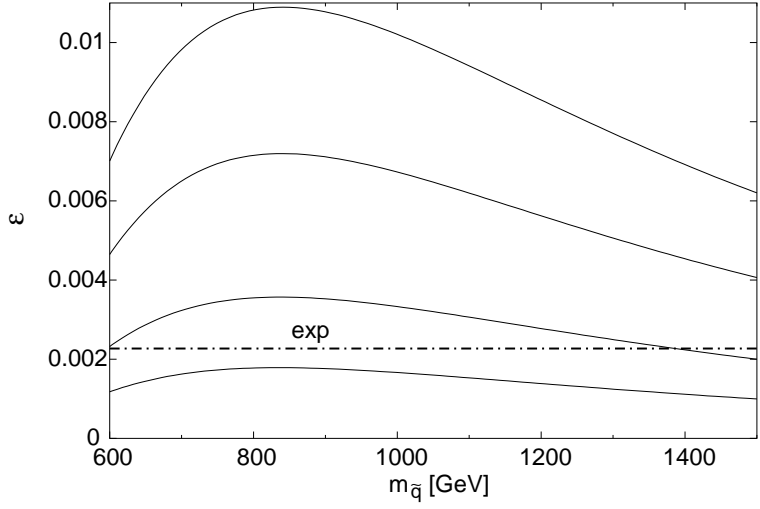,width=10cm}
\end{center}
\caption{$\epsilon^{\rm SUSY}$ in the large angle MSW case for
  $M_{\nu_R} = 5 \times 10^{13}$, $1 \times 10^{14}$, $2 \times
  10^{14}$ and $3 \times 10^{14} \rm {GeV}$ from the below.  We take
  $M_2 = 150 ~{\rm GeV}$, $a_0 = 0$, $\tan \beta =3$.  The horizontal
  dash-dotted line corresponds to experimental value $\epsilon^{\rm
    exp} = 2.271 \times 10^{-3}$. }
\label{fig2}
\end{figure}

In Fig. \ref{fig2} we show the dependence of $\epsilon^{{\rm SUSY}}$
on squark mass $m_{\tilde{q}}$ in the large angle MSW case, for
$M_{\nu_R} = 5\times 10^{13}$, $1 \times 10^{14}$, $ 2 \times 10^{14}$
and $3 \times 10^{14}$ GeV.  As one can see from
Eq.(\ref{eq:rough_epsilon}), the higher the right-handed neutrino mass
scale is, the larger $\epsilon^{{\rm SUSY}}$ becomes.  If $M_{\nu_R}
\gtrsim 10^{14}$ GeV, $\epsilon^{{\rm SUSY}}$ is comparable to the
experimental value as estimated before.  It is easy to understand the
dependence on $m_{\tilde{q}}$.  When $m_0$ is large, $\Delta^{(R)}$ is
also enhanced.  On the other hand when squarks become much heavier,
they decouple from low energy physics and $\epsilon^{{\rm SUSY}}$ gets
smaller.  Hence, for $M_2 = 150 \ {\rm GeV}$ corresponding to $M_3
\simeq 450 \ {\rm GeV}$, $\epsilon^{{\rm SUSY}}$ takes the maximum
value at $m_{\tilde{q}} \simeq 800$ GeV and monotonously decreases for
$m_{\tilde{q}} \gtrsim 800$ GeV.

\begin{figure}
\begin{center}
\leavevmode 
\psfig{file=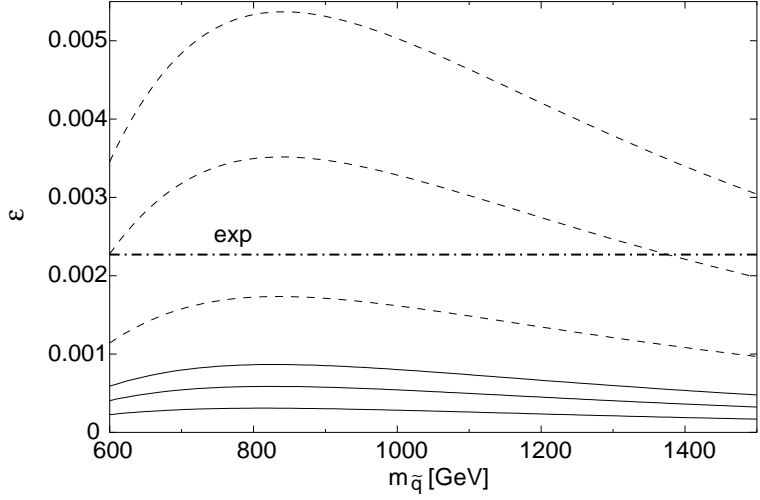,width=10cm}
\end{center}
\caption{$\epsilon^{\rm SUSY}$ in the small angle MSW case, for 
  $s^{(L)}_{13} = 0$ (solid line) and $s^{(L)}_{13} = 0.02$ (dashed line). 
  In both cases, we choose $M_{\nu_R} = 1 \times 10^{14}$, 
  $2 \times 10^{14}$ and $3 \times 10^{14} ~{\rm GeV}$ 
  from the below. We take $M_2 = 150 ~{\rm GeV}$, $a_0 = 0$ and 
  $\tan \beta = 3$. 
  The horizontal dash-dotted line corresponds to experimental value 
  $\epsilon^{\rm exp} = 2.271 \times 10^{-3}$. }
\label{fig3}
\end{figure}

The same dependence in small angle MSW solution is shown in Fig
\ref{fig3}, for $M_{\nu_R} = 1 \times 10^{14}$, $2 \times 10^{14}$ and
$3 \times 10^{14} ~{\rm GeV}$.  Solid lines correspond to
$s^{(L)}_{13} = 0$ case.  In this case $\epsilon^{\rm SUSY}$ is much
smaller than that in large angle solution since $s^{(L)}_{12}$ is very
small.  However if $s^{(L)}_{13}$ is nonzero, the situation changes
drastically.  Dashed lines show results with $s^{(L)}_{13} = 0.02$.
In this case $\epsilon^{\rm SUSY}$ is enhanced and can be comparable
to $\epsilon^{\rm exp}$ since $m_{\nu_3}$ is larger than $m_{\nu_2}$
by one order of magnitude and $s^{(L)}_{32}$ is also large.  As
$s^{(L)}_{13}$ gets close to the current upper limit of about 0.15,
$\epsilon^{\rm SUSY}$ becomes much larger.\footnote{ The effect of
  finite $s^{(L)}_{13}$ on $\mu \to e \gamma$ has been discussed in
  \cite{HisanoNomura}.} Hence improvement of the limit on
$s^{(L)}_{13}$ obtained by reactor experiments is very important to
discuss CP violation in the kaon system in our model, especially for
small angle case.

It is important to consider the implication of the SUSY contribution
to $\epsilon$ to the KM matrix since in the standard model, the
$\epsilon$ parameter provides an important constraint on the $\rho$
vs.\ $\eta$ plane.  In the standard model, the $\epsilon$ parameter is
given as (see, for example, \cite{hph9704376})
\begin{eqnarray}
    \epsilon^{\rm SM} \simeq C_\epsilon B_K A^2 \lambda^6 \eta
    \left[ - \eta_1 S_0 (x_c) + \eta_3 S_0 (x_c,x_t)
        + A^2 \lambda^4 (1-\rho) \eta_2 S_0 (x_t)
    \right] e^{i\pi/4},
    \label{eps_sm}
\end{eqnarray}
where $C_\epsilon=3.78\times 10^{4}$, $\eta_1$ $-$ $\eta_3$ are
parameters for the QCD factors given by $\eta_1=1.38\pm 0.20$,
$\eta_2=0.57\pm 0.01$, and $\eta_3=0.47\pm 0.04$, and Inami-Lim
functions \cite{inami81} $S_0(x_{c,t})$ and $S_0(x_c,x_t)$ are given by 
\begin{eqnarray}
    S_0 (x_{c,t}) &=& 
    \frac{4x_{c,t}-11x_{c,t}^2+x_{c,t}^3}{4(1-x_{c,t})^2}
    - \frac{3x_{c,t}^3\log x_{c,t}}{2(1-x_{c,t})^3},
    \\
    S_0 (x_c,x_t) &=& x_c \left[ \log \left(\frac{x_t}{x_c}\right) 
        - \frac{3x_t}{4(1-x_t)} - \frac{3x_t^2\log x_t}{4(1-x_t)^2}
    \right],
\end{eqnarray}
with $x_{c,t}=m_{c,t}^2/m_W^2$. In addition, we use $B_K=0.75-1.10$
\cite{PDG}.  Comparing Eq.\ (\ref{eps_sm}) with the experimentally
measured value of $\epsilon$, a constraint on the $\rho$ vs.\ $\eta$
plane is derived in the standard model.

If SUSY contribution to $\epsilon$ exists, such a constraint should be
reconsidered.  In particular, as we have seen, $\epsilon^{\rm SUSY}$
may be as large as the experimetally measured value of $\epsilon$, and
hence the observed value of $\epsilon$ may have a significant amount
of a contamination of $\epsilon^{\rm SUSY}$.  Since the size of
$\epsilon^{\rm SUSY}$ is model dependent and hence is unknown, we
should disregard the constraint from the $\epsilon$ parameter. Of
course, the $\rho$ vs.\ $\eta$ plane is still constrained by other
quantities like $|[V_{\rm KM}]_{ub}/[V_{\rm KM}]_{cb}|$ and $\Delta
m_{B_d}$, which are related to the parameters in the KM matrix as
$|[V_{\rm KM}]_{ub}/[V_{\rm KM}]_{cb}|\simeq\lambda
(\rho^2+\eta^2)^{1/2}$, and 
\begin{eqnarray}
    \Delta m_{B_d} \simeq 
    \frac{G_{\rm F}^2}{6\pi^2} \eta_B m_{B_d} (F_{B_d}^2B_{B_d})
    m_W^2 S_0 (x_t) A^6 \lambda^2 \left[ (1-\rho)^2 + \eta^2 \right],
\label{dmb_sm}
\end{eqnarray}
where $F_{B_d}\sqrt{B_{B_d}}=(200\pm 40)\ {\rm MeV}$, and
$\eta_B=0.55\pm 0.01$ \cite{hph9704376}.

Importantly, the allowed region on the $\rho$ vs.\ $\eta$ plane
changes as we exclude the constraint from $\epsilon$.  Accordingly,
possible size of the CP violation in the KM matrix, i.e., the
$\eta$-parameter, changes.

To make a more quiantitative discussion, we derived constraints on the
$\rho$ vs.\ $\eta$ plane comparing the standard model predictions on
$\epsilon$, $|[V_{\rm KM}]_{ub}/[V_{\rm KM}]_{cb}|$ and $\Delta
m_{B_d}$ with their experimental values.\footnote{Upper bound on
  $\Delta m_{B_s}$ also provides another constraint.  Inclusion of
  such information, however, does not change the following result
  significantly, and hence we neglect it in our analysis.} With and
without the constraint from $\epsilon$, we found the upper and lower
bounds on the $\eta$ parameter become\footnote{For the case without
  $\epsilon$, negative value of $\eta$ is also allowed.  However,
  $\eta<0$ is strongly disfavored by the BELLE and BABAR experiments
  \cite{BELLE,BABAR}, and we do not consider such a case.}
\begin{eqnarray}
    0.26 \leq \eta \leq 0.48 &:& \mbox{with $\epsilon$},
    \\
    0.17 \leq \eta \leq 0.53 &:& \mbox{without $\epsilon$},
\end{eqnarray}
where we assumed flat distributions of the uncertainties in various
parameters given above, and we adopted $A=0.834\pm 0.039$ \cite{PDG}.
As one can see, the upper and lower bounds on $\eta$ changes by about
10 \% and 35 \%, respectively, if the constraint from $\epsilon$ is
not taken into account.  This fact has important implications to
future studies of CP violations using rare processes like
$K_L\rightarrow\pi^0\nu\bar{\nu}$, as will be discussed in the
following subsection.

\subsection{$K_L\rightarrow\pi^0\nu\bar{\nu}$}

In the future, we may have another interesting information of the CP
violation via the rare process $K_L\rightarrow\pi^0\nu\bar{\nu}$.  In
the standard model, this process is mostly a signal of direct CP
violation, and the amplitude for this mode is proportional to $\eta$.
Most importantly, hadronic contribution to
$K_L\rightarrow\pi^0\nu\bar{\nu}$ is small, and short distance effects
of QCD are well under control \cite{buras96}. Therefore, this mode is
theoretically very clean.  Although the branching ratio for this
process is very small, we may have substantial number of
$K_L\rightarrow\pi^0\nu\bar{\nu}$ event in future experiments
\cite{KOPIO}, which may provide important test of the unitarity of the
KM matrix.  Due to Ref.\ \cite{KOPIO}, the $\eta$ parameter may be
determined at 10 \% accuracy.

This fact has an important implication to the search for the new
physics.  In the standard model, the $\eta$ parameter determined by
$K_L\rightarrow\pi^0\nu\bar{\nu}$ should be consistent with that from
other constraints, in particular, that from $\epsilon$.  If the SUSY
contribution to $\epsilon$ is sizable, however, this may not be
realized.  Indeed, as discussed in the previous subsection, the upper
and lower bounds on $\eta$ may vary by about 10 \% and 35 \%,
respectively, if the constraint from $\epsilon$ is not taken into
account. Given the fact that $\eta$ may be determined with the
accuracy of 10 \% \cite{KOPIO}, measurement of $\eta$ in the
experiment of $K_L\rightarrow\pi^0\nu\bar{\nu}$ gives an important
constraint on $\rho$ vs.\ $\eta$ plane independetly with the
measurement of $\epsilon$.

Of course, the Wilson coefficients for
$K_L\rightarrow\pi^0\nu\bar{\nu}$ may be directly affected by the SUSY
loops. Thus, we also calculated the SUSY correction to the
$K_L\rightarrow\pi^0\nu\bar{\nu}$ process in our framework.

There are penguin and box diagrams in SUSY loop corrections which
induce $K_L\rightarrow\pi^0\nu\bar{\nu}$. Effective operators are
${\cal Q}_{LL}^\nu \equiv (\bar{d}_L \gamma^\mu s_L)(\bar{\nu}_L\gamma_\nu
\nu_L)$ and ${\cal Q}_{RL}^\nu \equiv (\bar{d}_R \gamma^\mu
s_R)(\bar{\nu}_L\gamma_\nu \nu_L)$ with corresponding Wilson
coefficients $C_{LL}^{\nu}$ and $C_{RL}^\nu$.  SUSY contributions to
Wilson coefficients are summarized in the Appendix \ref{app:kpnn}.
For the process $K_L\rightarrow\pi^0\nu\bar{\nu}$, we found that the
total SUSY contribution to the Wilson coefficients is a few \% of the
SM contribution, and that the SUSY contribution has distractive
interference with the SM one in our model.  As a result, it is
challenging to see this deviation in the experiments.

For reasonable choices of parameters, the dominant SUSY contributions are
from chargino and charged higgs penguin diagrams.  Thus, the CP
violating phase is controled by the phase in the KM matrix since the
SUSY contribution is approximatelhy proportional to the 1-2 element of
the left-handed squark mass matrix.  On the contrary, effects of the
GUT phases are tiny in $K_L\rightarrow\pi^0\nu\bar{\nu}$ because of
the smalless of neutralino contribution which is, we found, mostly 0.1
\% level.  Since the SUSY contributions to this mode are quite small to
be observed, useful information about the parameter $\eta$ is expected
from $Br(K_L\rightarrow\pi^0\nu\bar{\nu})$ in our model.

\subsection{$\epsilon' / \epsilon$}

In supersymmetric models, the parameter $\epsilon'$ can be also
modified due to the supersymmetric loop effects.  In particular, in
Ref.\ \cite{PRL83-4929}, it was pointed out that the SUSY contribution
to $\epsilon'/\epsilon$ may become much larger than the experimental
value, $(\epsilon'/\epsilon)^{\rm exp}=(19.3 \pm 2.4)\times 10^{-4}$ \cite{KTeVNA48}, because
of the significant modification of the $\Delta I=\frac{3}{2}$
amplitude contributing to $\epsilon'/\epsilon$.\footnote{In SUSY
models, $\epsilon'/\epsilon$ can be also modified if the
$A$-parameters for the down-type quarks are not aligned to the
down-type Yukawa matrix \cite{PRL83-907}.} As a result, in some
parameter space, $\epsilon'/\epsilon$ provides severer constraint on
CP and flavor violations in the soft SUSY breaking parameters than the
$\epsilon$ does, and hence one might worry if the SUSY contribution to
$\epsilon'/\epsilon$ is in a reasonable range in our framework.  In
this subsection, we discuss the SUSY contribution to
$\epsilon'/\epsilon$.

Before discussing the supersymmetric effect on $\epsilon'/\epsilon$,
let us first introduce several formulae necessary to evaluate
$\epsilon'/\epsilon$.  The $\epsilon'/\epsilon$ parameter is
calculated once the Wilson coefficients for the $\Delta S=1$ effective
Hamiltonian are given.  The $\Delta S=1$ effective Hamiltonian is
denoted as\footnote{We use the suffix $(\Delta S=1)$ for the Wilson
coefficients and operators for the $\Delta S=1$ effective Hamiltonian
to distinguish them from those for the $\Delta S=2$ effective
Hamiltonian.}
\begin{eqnarray}
    {\cal H}_{\rm eff}^{(\Delta S=1)} &=& 
    \sum_i \left( 
        C_{L,i}^{(\Delta S=1)} {\cal Q}_{L,i}^{(\Delta S=1)}
        + C_{R,i}^{(\Delta S=1)} {\cal Q}_{R,i}^{(\Delta S=1)}
    \right),
    \label{H_epspr}
\end{eqnarray}
where $C_i$ are Wilson coefficients, and 
\begin{eqnarray}
    {\cal Q}_{L,3}^{(\Delta S=1)}
    &=& 4 (\bar{s}_\alpha \gamma_\mu P_L d_\alpha)
    \sum_q (\bar{q}_\beta \gamma_\mu P_L q_\beta),
    \\
    {\cal Q}_{L,4}^{(\Delta S=1)}
    &=& 4 (\bar{s}_\alpha \gamma_\mu P_L d_\beta)
    \sum_q (\bar{q}_\beta \gamma_\mu P_L q_\alpha),
    \\
    {\cal Q}_{L,5}^{(\Delta S=1)}
    &=& 4 (\bar{s}_\alpha \gamma_\mu P_L d_\alpha)
    \sum_q (\bar{q}_\beta \gamma_\mu P_R q_\beta),
    \\
    {\cal Q}_{L,6}^{(\Delta S=1)}
    &=& 4 (\bar{s}_\alpha \gamma_\mu P_L d_\beta)
    \sum_q (\bar{q}_\beta \gamma_\mu P_R q_\alpha),
    \\
    {\cal Q}_{L,7}^{(\Delta S=1)}
    &=& 4 (\bar{s}_\alpha \gamma_\mu P_L d_\alpha)
    \sum_q e_q (\bar{q}_\beta \gamma_\mu P_R q_\beta),
    \\
    {\cal Q}_{L,8}^{(\Delta S=1)} 
    &=& 4 (\bar{s}_\alpha \gamma_\mu P_L d_\beta)
    \sum_q e_q (\bar{q}_\beta \gamma_\mu P_R q_\beta),
    \\
    {\cal Q}_{L,9}^{(\Delta S=1)} 
    &=& 4 (\bar{s}_\alpha \gamma_\mu P_L d_\alpha)
    \sum_q e_q (\bar{q}_\beta \gamma_\mu P_L q_\beta),
    \\
    {\cal Q}_{L,10}^{(\Delta S=1)} 
    &=& 4 (\bar{s}_\alpha \gamma_\mu P_L d_\beta)
    \sum_q e_q (\bar{q}_\beta \gamma_\mu P_L q_\alpha),
\end{eqnarray}
with $P_{L/R}=\frac{1}{2}(1\mp\gamma_5)$, $\alpha$ and $\beta$ being
the color indices, $e_q$ denoting the quark electric charge, and
${\cal Q}_{R,i}^{(\Delta S=1)}\equiv {\cal Q}_{L,i}^{(\Delta
S=1)}|_{L\leftrightarrow R}$.  Based on the structure of the
operators, we call ${\cal Q}_{L,i}^{(\Delta S=1)}$ with $i=3$, 4, 9
and 10 (${\cal Q}_{R,i}^{(\Delta S=1)}$ with $i=3$, 4, 9 and 10) as
$LL$-type ($RR$-type) operators while others as $LR$- or $RL$-type
operators.

In the standard model, the gluon-penguin diagram generates the
operators ${\cal Q}_{L,i}^{(\Delta S=1)}$ with $i=3-6$, and hence
their Wilson coefficients are of $O(\alpha_{\rm W}\alpha_{\rm s})$
where $\alpha_{\rm W}$ indicates the coupling constants for the
electroweak interaction.  On the contrary, the operators ${\cal
Q}_{L,i}^{(\Delta S=1)}$ with $i=7-10$ are only from the electroweak
processes and the corresponding Wilson coefficients are of
$O(\alpha_{\rm W}^2)$.  Although the Wilson coefficients contributing
to the $\Delta I=\frac{3}{2}$ amplitude are smaller than those
contributing to the $\Delta I=\frac{1}{2}$ one, both amplitudes are
significant since the $\Delta I=\frac{3}{2}$ contribution has extra
enhancement factor.

From the effective Hamiltonian given in Eq.\ (\ref{H_epspr}),
$\epsilon'/\epsilon$ is given by (see, for example, \cite{hph9704376})
\begin{eqnarray}
    \frac{\epsilon'}{\epsilon} &=& 
    \frac{G_{\rm F}\omega}{2|\epsilon|{\rm Re}A_0}
    \sum_i {\rm Im} C_{L,i}^{(\Delta S=1)}
    \left[ (1-\Omega_{\eta\eta'})
        \langle (\pi\pi)_{I=0} | {\cal Q}_{L,i}^{(\Delta S=1)}
        | K \rangle
        - \frac{1}{\omega} \langle (\pi\pi)_{I=2} | 
        {\cal Q}_{L,i}^{(\Delta S=1)} | K \rangle
    \right]
    \nonumber \\ &&
    + (L\rightarrow R),
    \label{e'/e}
\end{eqnarray}
where $|(\pi\pi)_{I}\rangle$ denotes the two-pion state with isospin
$I$, and
\begin{eqnarray}
\omega \equiv \frac{{\rm Re}A_2}{{\rm Re}A_0},
\end{eqnarray}
with
\begin{eqnarray}
    A_0 \equiv 
    \langle (\pi\pi)_{I=0} | {\cal H}_{\rm eff} | K \rangle,
    ~~~
    A_2 \equiv 
    \langle (\pi\pi)_{I=2} | {\cal H}_{\rm eff} | K \rangle.
\end{eqnarray}
Numerically, $\omega\simeq 0.045$.  In addition, $\Omega_{\eta\eta'}$
is from the isospin breaking in the quark masses, and numerically we
use $\Omega_{\eta\eta'}= 0.25$ \cite{hph9704376}.  Since $\omega$ is a
small number, the $\Delta I=\frac{3}{2}$ contribution in Eq.\ 
(\ref{e'/e}), which is proportional to $\omega^{-1}$, is enhanced
relative to the $\Delta I=\frac{1}{2}$ contribution.

Now, let us consider the supersymmetric contribution to
$\epsilon'/\epsilon$.  In MSSM, the Wilson coefficients for the
$\Delta S=1$ effective Lagrangian are modified by integrating out
supersymmetric particles.  (Formulae for the SUSY contribution to the
Wilson coefficients are given in Appendix \ref{app:dS=1}.)  With the
SUSY contribution to the Wilson Coefficients given at the electroweak
scale, we can calculate $\epsilon'/\epsilon$ using the same
prescription as the SM case.

One important observation is that, in MSSM, $\Delta I=\frac{3}{2}$
contribution can be largely enhanced relative to the standard model
case \cite{PRL83-4929}.  This is because the $\Delta I=\frac{3}{2}$
contribution is, as mentioned before, of $O(\alpha_{\rm W}^2)$ in the
standard model while it can be of $O(\alpha_{\rm s}^2)$ in MSSM if the
mass matrix of the down-type squarks has non-vanishing 1-2 element.
Combining the fact that $\Delta I=\frac{3}{2}$ contribution is
enhanced by the factor $\omega^{-1}\simeq 22$, the SUSY contribution
to the $\epsilon'/\epsilon$ parameter can be order of magnitude larger
than the experimental value.  Thus, in some case, constraint on the CP
and flavor violating parameters in MSSM from $\epsilon'/\epsilon$
becomes severer than that from $\epsilon$.

As we will see below, the SUSY contribution to $\epsilon'/\epsilon$ is
relatively small in our framework.  Before showing the numerical
results, let us explain how the the smallness of $\epsilon'/\epsilon$
is understood in our framework compared to the result given in Ref.\ 
\cite{PRL83-4929}.

As a first step, let us briefly review how the large SUSY contribution
is realized in Ref.\ \cite{PRL83-4929}.  The most important
enhancement is for the the $LR$-type $\Delta I=\frac{3}{2}$ amplitude
by the factor of $\sim\alpha_{\rm s}^2/\alpha_{\rm W}^2$ relative to
the SM contributions.  The same enhancement also exists for the $LL$-
and $RR$-type operators.  However, in taking the relevant matrix
elements of the $LR$-type operators, there is a chirality-enhancement
factor proportional to $m_K^2/m_s^2\sim 10$, and hence the $LR$-type
operators are more important than the $LL$- and $RR$ ones.

Since the $\Delta I=\frac{3}{2}$ amplitude is an isospin-breaking
effect, a mass difference between the up- and down-squark masses is
necessary.  For the right-handed up- and down-squarks, significant
mass splitting is realized if the SUSY breaking mass parameters for
them take different values.  On the contrary, the left-handed ones are
both from the SU(2)$_L$-doublet.  As a result, their mass splitting is
only from the electroweak symmetry breaking effect (i.e., vacuum
expectation values of the Higgs bosons), and the mass splitting
between left-handed squarks is too small to generate significant
contribution to the isospin-breaking amplitude.  Thus, in order to
generate $LR$-type operators, sizable 1-2 element of the left-handed
down-type squark mass matrix is necessary.  In summary, the condition
for the large SUSY contribution to $\epsilon'/\epsilon$ is (i) mass
splitting of between $\tilde{u}_R$ and $\tilde{d}_R$, and (ii)
non-vanishing imaginary part of $[m_{\tilde{d}_L}^2]_{12}$.  Combining
all of these effects, the SUSY contribution to $\epsilon'/\epsilon$
can be as large as $10^{-2}$ when ${\rm
  Im}([m_{\tilde{d}_L}^2]_{12}/m^2_{\tilde{d}})\sim 10^{-2}$.

Now, we turn to our case.  In our framework, large off-diagonal
elements are generated for the right-handed down-type squark mass
matrix since they are related to the neutrino Yukawa interactions.
Since sizable mass splitting between up- and down-type squarks is
possible only for the right-handed sector, only the $RR$-type
contribution is modified for the $\Delta I=\frac{3}{2}$ process, and
hence the chirality enhancement is not significant for $\Delta
I=\frac{3}{2}$ process.  In addition, if we take the universal
boundary condition for the squark masses, mass splitting between
$\tilde{u}_R$ and $\tilde{d}_R$ is fairly small.  As a result,
correction to the $\Delta I=\frac{3}{2}$ amplitude becomes small.

To make more quantitative discussion, we calculate the SUSY
contributions to the $\Delta I=\frac{1}{2}$ and $\Delta I=\frac{3}{2}$
part of $\epsilon'/\epsilon$.  In our analysis, we followed the
prescription given in Ref.\ \cite{hph9704376} to calculate the SUSY
contribution to $\epsilon'/\epsilon$.  We first calculate the SUSY
contribution to the Wilson coefficients at the electroweak scale.
Then, we run them down to the charm quark mass scale using the
relevant RGEs and took the matrix elements.  The resultant SUSY
contribution is linear in the Wilson coefficients, and numerically we
found
\begin{eqnarray}
    (\epsilon' / \epsilon)^{\rm SUSY} =
    \sum_i N_i {\rm Im} 
    \left[
        \tilde{C}_{L,i}^{(\Delta S=1)} (\mu =m_W)
        + \tilde{C}_{R,i}^{(\Delta S=1)} (\mu =m_W)
    \right],
\end{eqnarray}
where $\tilde{C}_{L,i}^{(\Delta S=1)}$ and $\tilde{C}_{R,i}^{(\Delta
S=1)}$ are SUSY contribution to the corresponding Wilson coefficients,
and the numerical values for $N_i$ are given in Table \ref{table:Ni}.

\begin{table}[t]
    \begin{center}
        \begin{tabular}{rrrr}
            \hline\hline
            {$m_s$} & {0.11 GeV} & {0.13 GeV} & {0.15 GeV} \\
            \hline
            {$N_3$} & 
            {$6.64\times 10^{4}$} &
            {$8.92\times 10^{5}$} &
            {$1.42\times 10^{6}$} \\
            {$N_4$} & 
            {$-9.98\times 10^{6}$} &
            {$-9.01\times 10^{6}$} &
            {$-8.39\times 10^{6}$} \\
            {$N_5$} & 
            {$1.44\times 10^{7}$} &
            {$1.07\times 10^{7}$} &
            {$8.29\times 10^{6}$} \\
            {$N_6$} & 
            {$4.30\times 10^{7}$} &
            {$3.22\times 10^{7}$} &
            {$2.52\times 10^{7}$} \\
            {$N_7$} & 
            {$8.28\times 10^{8}$} &
            {$5.97\times 10^{8}$} &
            {$4.49\times 10^{8}$} \\
            {$N_8$} & 
            {$2.60\times 10^{9}$} &
            {$1.89\times 10^{9}$} &
            {$1.43\times 10^{9}$} \\
            {$N_9$} & 
            {$9.08\times 10^{6}$} &
            {$1.10\times 10^{7}$} &
            {$1.23\times 10^{7}$} \\
            {$N_{10}$} & 
            {$2.32\times 10^{6}$} &
            {$3.46\times 10^{6}$} &
            {$4.19\times 10^{6}$} \\
            \hline\hline
        \end{tabular}
        \caption{Coefficients $N_i$ in units of GeV$^2$ for the
        fitting formula of the $\epsilon'/\epsilon$ parameter. We use
        the bag parameters given in Ref.\ \cite{hph9704376}, and
        $m_s=0.11$, 0.13, and 0.15 GeV.}
\label{table:Ni} 
\end{center}
\end{table}

\begin{figure}
\begin{center}
\leavevmode 
\psfig{file=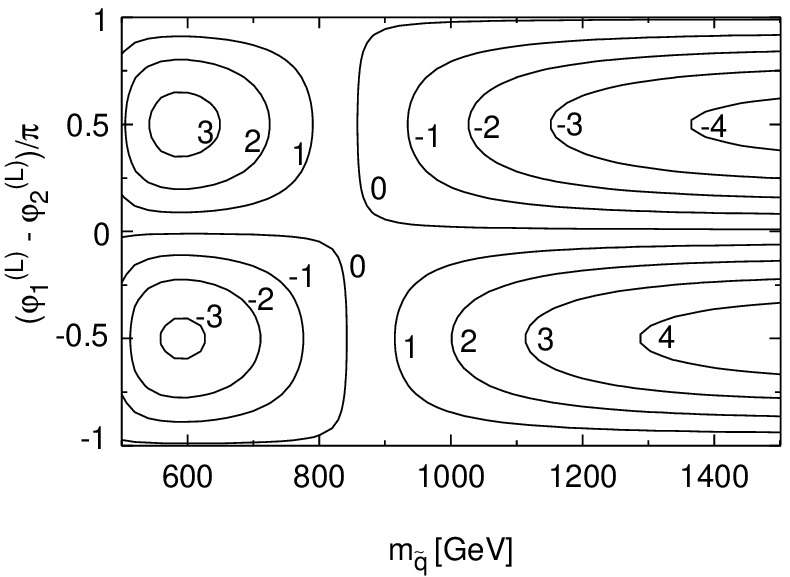,width=10cm}
\end{center}
\caption{Contours of constant $R_{1/2}$ in units of $10^{-2}$, which is
  $(\epsilon'/\epsilon)^{\rm SUSY}_{\Delta I=1/2}$ normalized by the
  standard model contribution.  Here, we consider the large angle MSW
  case and take $M_{\nu_R}=3\times 10^{14}\ {\rm GeV}$, $\tan\beta=3$,
  and $m_s=0.13$ GeV.}
\label{fig:epp1/2}
\begin{center}
\leavevmode 
\psfig{file=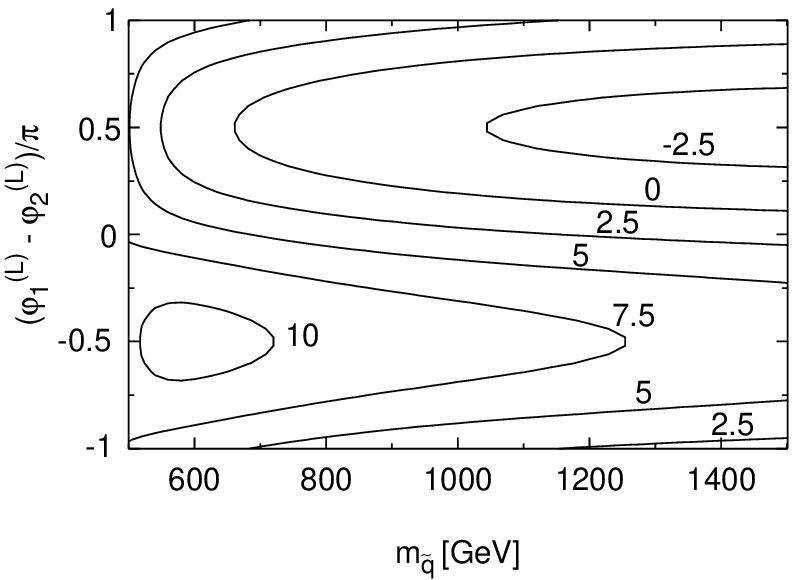,width=10cm}
\end{center}
\caption{Contours of constant $R_{3/2}$ in units of $10^{-2}$, which is
  $(\epsilon'/\epsilon)^{\rm SUSY}_{\Delta I=3/2}$ normalized by the
  standard model contribution.  Here, we consider the large angle MSW
  case and take $M_{\nu_R}=3\times 10^{14}\ {\rm GeV}$, $\tan\beta=3$,
  and $m_s=0.13$ GeV.}
\label{fig:epp3/2}
\end{figure}

First, we present the SUSY contribution to the $\Delta I=\frac{1}{2}$
and $\Delta I=\frac{3}{2}$ part of $\epsilon'/\epsilon$ normalized by
the standard model ones:
\begin{eqnarray}
    R_{1/2} \equiv 
    \frac{(\epsilon'/\epsilon)^{\rm SUSY}_{\Delta I=1/2}}
    {(\epsilon'/\epsilon)^{\rm SM}_{\Delta I=1/2}},~~~
    R_{3/2} \equiv 
    \frac{(\epsilon'/\epsilon)^{\rm SUSY}_{\Delta I=3/2}}
    {(\epsilon'/\epsilon)^{\rm SM}_{\Delta I=3/2}}.
\end{eqnarray}
In Figs.\ \ref{fig:epp1/2} and \ref{fig:epp3/2}, we plot contours of
constant $R_{1/2}$ and $R_{3/2}$ on $m_{\tilde{q}}$ vs.
$\varphi^{(L)}_1-\varphi^{(L)}_2$ plane for the case of the large
angle MSW.  Here, we take $M_{\nu_R}=3\times 10^{14}\ {\rm GeV}$, and
$\tan\beta=3$.  As one can see, the SUSY contribution is at most a few
\% of the standard model contribution for the $\Delta I=\frac{1}{2}$
part, and $\sim 10\ \%$ for the $\Delta I=\frac{3}{2}$ one.  It is
notable that, in our simple analysis, universal scalar mass is assumed
as a boundary condition.  Consequently, mass splitting between the
right-handed up- and down-type squarks becomes small since the masses
of these squarks are mostly determined by the boundary condition and
the RG effect due to the gluino loop below the GUT scale which are
universal in this case.  As a result, masses of the right-handed up-
and down-type squarks are quite degenerate, and this fact gives
another suppression factor for the $\Delta I=\frac{3}{2}$
contribution.  Since the contributions to the $\Delta I=\frac{1}{2}$
and $\Delta I=\frac{3}{2}$ amplitudes are both small, the SUSY
contribution to the $\epsilon'/\epsilon$ parameter is also at most a
few \% level in our framework.

\section{Conclusions and Discussion}
\label{sec:conclusion}
\setcounter{equation}{0}

In this paper, we discussed CP violations in the supersymmetric SU(5)
model with right-handed neutrinos.  In this class of model,
off-diagonal elements of the right-handed down-type squarks are
generated via the RG effect even if they vanish at
the cut-off scale (i.e., in our case, the reduced Planck scale).  In
general, such off-diagonal elements contain CP violating phases and
they can be extra sources of the CP violations in the low-energy
processes.  In particular, it was emphasized that such phases can be
related to phases in the unified theories, which are not related to
the parameters in the standard model.

We paid particular attentions to the CP violations in the kaon system.
Most importantly, we have seen that the $\epsilon$ parameter can be
severely affected by the SUSY contribution; even with a relatively
heavy squark mass of $\sim 1\ {\rm TeV}$, the SUSY contribution to
$\epsilon$ can be as large as the experimentally measured value if the
Majorana mass for the right-handed neutrinos is as large as $\sim
10^{14}\ {\rm GeV}$.  Of course, the SUSY contribution to $\epsilon$
strongly depends on the phases in the off-diagonal elements of the
squark mass matrix.  As we have seen, the phases in these parameters
can be naturally large due to the phases in the GUT model.

We have also calculated the SUSY contribution to
$Br(K_L\rightarrow\pi^0\nu\bar{\nu})$ and $\epsilon'/\epsilon$.
Unfortunately, however, the SUSY contributions to these quantities are
relatively small.  In our framework, SUSY contribution to
$Br(K_L\rightarrow\pi^0\nu\bar{\nu})$ is a few \%, and the SUSY
contribution to $\epsilon'/\epsilon$ is also a few \% level of the
standard model one.

Thus, in our framework, the SUSY contribution is the most important
for the $\epsilon$ parameter.  This fact has an important implication
to the future test of the unitarity of the KM matrix, since $\epsilon$
provides one of the important information of the magnitude of the CP
violation in the KM matrix in the standard model.  In the standard
model, the so-called $\rho$ vs.\ $\eta$ plane was constrained so that
the standard-model prediction of the $\epsilon$ parameter agrees with
observed one.  If the SUSY contribution to $\epsilon$ is sizable,
however, $\epsilon$ cannot be used to constrain the $\rho$ vs.\ $\eta$
plane, which may change the bounds on $\rho$ and $\eta$.  We have seen
that the upper and lower bounds on the $\eta$ parameter changes by
about 10 \% and 35 \%, respectively, if we discard the constraint from
$\eta$.  The deviation from the standard-model prediction on $\eta$
may be tested by future experiments.  In particular, the measurements
of $\phi_1$ and $Br(K_L\rightarrow\pi^0\nu\bar{\nu})$ will provide an
interesting test of the value of $\eta$.  Notice that there are still
sizable uncertainties in the theoretical calculation of the $\epsilon$
parameter, in particular from the bag parameter in taking the hadronic
matrix elements and from the Wolfenstein's $A$-parameter in the KM
matrix.  Thus, reduction of the uncertainties in these quantities will
be very important to find a signal of the SUSY loop using the
$\epsilon$ parameter.

{\sl Note Added:} In finalizing this paper, we found a paper by
S. Baek et al. \cite{baek01} which has some overlap with our analysis.
In particular, in \cite{baek01}, authors paid special attention to
non-minimal contribution to the Yukawa matrices at the GUT scale in
order to realize a realistic unification of the down-type and
charged-lepton Yukawa matrices.

{\sl Acknowledgment:} One of the authors (TM) would like to thank H.
Murayama for useful conversations.  The work of Y.K. is supported by
the Japan Society for the Promotion of Science. The work of T.M. is
supported by the Grant-in-aid from the Ministry of Education, Culture,
Sports, Science and Technology, Japan, No.12047201.

\appendix

\section{Interaction Lagrangian}
\setcounter{equation}{0}

In this section, we show the interaction Lagrangian used in this paper.
Relevant terms in our calculation are the vertices for charginos,
neutralinos and gluinos with squarks and quarks.  At first, we begin
relation between gauge and mass eigenstates and mixing matrix, which
make mass matrix diagonalized.
  
The mass and gauge eigenstates for squarks are denoted with
$\tilde{q_A}$ ($A=1,\cdots,6$), $\tilde{q_i}_{L/R}$ ($i=1,2,3$, label
for generation), respectively.  Then the relations between gauge and
mass eigenstates are given by:
\begin{eqnarray}
\tilde{q}_{i L} = [U(\tilde{q})]_{iA} \tilde{q}_A,~~~
\tilde{q}_{i R} &=& [U(\tilde{q})]_{(i+3)A} \tilde{q}_A,
\end{eqnarray}
where $U(\tilde{q})$ is a $6\times 6$ unitary matrix which diagonalize
squark mass matrix ${\cal M}^2_{\tilde{q}}$,
\begin{eqnarray}
[U^\dag (\tilde{q}) {\cal M}^2_{\tilde{q}} U(\tilde{q}) ]_{AB}
= m_{\tilde{q}_A}^2 \delta_{AB}.
\end{eqnarray}

Relation between gauge and mass eigenstate for charginos and
neutralinos are given by
\begin{eqnarray}
\left[ \begin{array}{c} 
       \tilde{W}_{L}^-\\
       \tilde{H}_{1L}^-
       \end{array}
\right]_{i} 
&=& 
[ U(\tilde{\chi}^-_{L}) ]_{iX} 
\tilde{\chi}^-_{X\,L},
\\
\left[ \begin{array}{c} 
       \tilde{W}_{R}^-\\
       \tilde{H}_{2R}^-
       \end{array}
\right]_{i} 
&=& 
[ U(\tilde{\chi}^-_{R}) ]_{iX} 
\tilde{\chi}^-_{X\,R},
\\
\left[ \begin{array}{c}
       \tilde{B}_{L}\\
       \tilde{W}_{L}^3\\
       \tilde{H}_{1L}^0\\
       \tilde{H}_{2L}^0\\
       \end{array}
\right]_{i}
&=&
[U(\tilde{\chi}^0)]_{iX} \tilde{\chi}^0_{X\,L},
\end{eqnarray}
where $U(\tilde{\chi}^-_{L/R})$ and $U(\tilde{\chi}^0)$ are $2\times2$
and $4\times 4$ unitary matrices which diagonalize mass matrix for
charginos and neutralinos, respectively.  Diagonalization of mass
matrices gives masses of charginos $m_{\tilde{\chi}^-_X}$ and
neutralinos $m_{\tilde{\chi}^0_X}$ ,
\begin{eqnarray}
[ U^\dag ( \tilde{\chi}^-_R ) 
  {\cal M}_{\tilde{\chi}^-} 
  U(\tilde{\chi}^-_L) ]_{XY}
= m_{\tilde{\chi}^-_X } \delta_{XY}, ~~~
[
U^\dag (\tilde{\chi}^0) {\cal M}_{\tilde{\chi}^0} U(\tilde{\chi}^0) 
]_{XY}
= m_{\tilde{\chi}^0_X} \delta_{XY}, 
\end{eqnarray}

For quarks, we chose the basis in which the mass eigenstates $(u,d)$
and gauge eigenstates $(u',d')$ are related as $d'_L=d_L$,
$u'_L=V_{\rm{KM}}~u_L$, $u'_R=u_R$ and $d'_R=d_R$.

The interaction Lagrangian for chargino-quark-squark couplings are
given by
\begin{eqnarray}
L_{{\tilde{\chi}^-}d\tilde{u}}
&=& \bar{\tilde{\chi}}^-_X 
\left( C_{dAX}^{L} P_L + C_{dAX}^{R} P_R \right) d
\tilde{u}_A^\ast 
\nonumber \\ &&
+
u^T (-C^\dag)
\left( C_{uAX}^{L} P_L + C_{uAX}^{R} P_R \right) 
\tilde{\chi}^-_X
\tilde{d}_A^\ast 
+ h.c., 
\label{vertex_C_Q_sQ_0}
\end{eqnarray}
where $C$ is the charge conjugation matrix,
$P_{L/R}=\frac{1}{2}\left(1\mp\gamma_5\right)$, and $u$ and $d$ denote
up-type ($u,c,t$) and down-type ($d,s,b$) quarks, respectively.
$C_{qAX}^{L,R}$ are given by
\begin{eqnarray}
C_{dAX}^L
&=& 
-g_2 [U^\ast (\tilde{\chi}^-_R) ]_{1X} 
[U^\ast (\tilde{u})]_{dA} 
+ 
[U^\ast (\tilde{\chi}^-_R)]_{2X} 
\sum_{j=1}^{3} [U^\ast (\tilde{u})]_{(3+j)A}
[\hat{Y}_U V_{\rm KM}]_{jd},
\label{vertex_C_Q_sQ_1}
\\
C_{dAX}^R 
&=& 
-
[U^\ast (\tilde{\chi}^-_L)]_{2X} 
[U^\ast (\tilde{u})]_{dA} [\hat{Y}_D]_{d},
\label{vertex_C_Q_sQ_2}
\\
C_{uAX}^L
&=& 
-g_2 [U (\tilde{\chi}^-_L)]_{1X} 
\sum_{j=1}^3 [U^\ast (\tilde{d}) ]_{jA} 
             [V_{\rm KM}^\ast ]_{uj}
\nonumber \\ &&
- 
[U(\tilde{\chi}^-_L)]_{2X} 
\sum_{j=1}^{3} [U^\ast (\tilde{d})]_{(3+j)A}
[\hat{Y}_D\, V_{KM}^\dag ]_{ju},
\label{vertex_C_Q_sQ_3}
\\
C_{uAX}^R 
&=& 
[U (\tilde{\chi}^-_R)]_{2X} 
\sum_{j=1}^3 [U^\ast (\tilde{d})]_{jA} 
[V_{\rm KM}^\dag\, \hat{Y}_U^\dag ]_{ju},
\label{vertex_C_Q_sQ_4}
\end{eqnarray}
where $g_2$ is the gauge coupling constant of SU(2)$_L$, and $u,
d=(1,2,3)$ for $(u,c,t)$ and $(d,s,b)$, respectively.

Neutralino-quark-squark coupling couplings are given by
\begin{eqnarray}
{\cal L}_{{\tilde{\chi}^0}d\tilde{d}}
&=&
\bar{\tilde{\chi}}^0_X 
\left( N_{dAX}^{L} P_L + N_{iAX}^{R} P_R \right) 
d \tilde{d}_A^\ast
\nonumber \\ && 
+
\bar{\tilde{\chi}}^0_X 
\left( N_{uAX}^{L} P_L + N_{uAX}^{R} P_R \right) 
u \tilde{u}_A^\ast + h.c. , 
\end{eqnarray}
where
\begin{eqnarray}
N_{dAX}^{L}
&=&
\left( - \frac{\sqrt{2}g_1}{6} 
[U(\tilde{\chi}^0) ]_{1X}
+\frac{\sqrt{2}g_2}{2} [U(\tilde{\chi}^0) ]_{2X}
\right)
[U^\ast (\tilde{d}) ]_{dA}
\nonumber \\ && 
+
[U(\tilde{\chi}^0) ]_{3X}
[U^\ast (\tilde{d}) ]_{(d+3)A}
[\hat{Y}_D ]_{d},
\\
N_{dAX}^{R}
&=&
- \frac{\sqrt{2}g_1}{3} 
[U^\ast(\tilde{\chi}^0) ]_{1X}
[U^\ast (\tilde{d}) ]_{(d+3)A}
+
[U^\ast(\tilde{\chi}^0) ]_{3X}
[U^\ast(\tilde{d}) ]_{dA}
[\hat{Y}_D^\dag ]_{d},
\\
N_{uAX}^{L}
&=&
\left( - \frac{\sqrt{2} g_1}{6} 
[U(\tilde{\chi}^0) ]_{1X}
-\frac{\sqrt{2}g_2}{2} [U(\tilde{\chi}^0) ]_{2X}
\right)
[U^\ast (\tilde{u}) ]_{uA}
\nonumber \\
&&
-
[U(\tilde{\chi}^0) ]_{4X}
\sum_{j=1}^3
[U^\ast (\tilde{u}) ]_{(j+3)A}
[\hat{Y}_U V_{\rm KM} ]_{ju},
\\
N_{uAX}^{R}
&=& 
\frac{2\sqrt{2} g_1}{3} 
[U^\ast(\tilde{\chi}^0) ]_{1X}
[U^\ast (\tilde{u}) ]_{(u+3)A}
\nonumber \\ &&
-
[U^\ast(\tilde{\chi}^0) ]_{4X}
\sum_{j=1}^3 
[U^\ast (\tilde{u}) ]_{jA}
[V_{\rm KM}^\dag \hat{Y}_U^\dag ]_{ju},
\end{eqnarray}
with $g_1$ being the gauge coupling constant of U(1)$_Y$.

The interaction Lagrangian for gluino-quark-squark coupling is given
by
\begin{eqnarray}
L_{\tilde{G}q\tilde{q}}
&=&
\tilde{d}_A^\ast \bar{\tilde{G}^a} T^a 
\left( G_{dA}^L P_L + G_{dA}^R P_R\right) d
\nonumber \\ && 
+
\tilde{u}_A^\ast \bar{\tilde{G}^a} T^a 
\left( G_{uA}^L P_L + G_{uA}^R P_R \right) u + h.c., 
\end{eqnarray}
where  
\begin{eqnarray}
G_{dA}^L &=& -\sqrt{2} g_s [U^\ast (\tilde{d})]_{iA},
\\
G_{dA}^R &=& \sqrt{2} g_s [U^\ast (\tilde{d})]_{(i+3)A},
\\
G_{uA}^L &=& -\sqrt{2} g_s 
   \sum_{j=1}^3
   [U^\ast (\tilde{u}) ]_{jA} 
   [V_{\rm KM}^\ast ]_{ju}, 
\\
G_{uA}^R &=& \sqrt{2} g_s [U^\ast (\tilde{u})]_{(u+3)A},
\end{eqnarray}
with $g_s$ being the gauge coupling constant of SU(3)$_C$.  Here we
take soft-breaking mass parameter of gauginos as real positive.

\section{SUSY Contribution to $K_L\rightarrow \pi^0 \nu \bar{\nu}$}
\label{app:kpnn}
\setcounter{equation}{0}

In this appendix, we show the SUSY contribution to the Wilson
coefficient of effective operators for
$K_L\rightarrow\pi^0\nu\bar{\nu}$.

There are sets of SUSY loop diagrams in which charged-Higgs, charginos
and neutralinos are exchanged, which induce effective operators
\begin{eqnarray}
{\cal Q}_{LL}^\nu \equiv 
(\bar{d} \gamma^\mu P_L s)
( \bar{\nu} \gamma^\mu P_L \nu ),~~~
{\cal Q}_{RL}^\nu \equiv 
(\bar{d} \gamma^\mu P_R s )
( \bar{\nu} \gamma^\mu P_L \nu ).
\end{eqnarray}
The Wilson coefficients of corresponding operators are referred to
$C_{LL}^\nu$ and $C_{RL}^\nu$, respectively.  The effective
Hamiltonian is given by ${\cal H}_{\rm eff}
=(C_{LL}^{ZP,\nu}+C_{LL}^{{\rm Box},\nu}) {\cal Q}_{LL}^\nu
+(C_{RL}^{ZP,\nu}+C_{RL}^{{\rm Box},\nu}) {\cal Q}_{RL}^\nu $.

In the following, each SUSY contribution of charged-Higgs, charginos and neutralinos are separately
shown for $Z$-penguin diagrams,
\begin{eqnarray}
\tilde{C}_{LL}^{ZP, \nu}|^{H^-}
&=&
\frac{g_Z^2}{2 M_Z^2}
m_u^2 \left(T^3_{u_R}-T^3_{u_L}\right)
{H^L_{ud}}^\ast H^L_{us}
I^0_3
\left(m_u^2,m_u^2,m_{H^-}^2 
\right)
\\
\tilde{C}_{LL}^{ZP, \nu}|^{\tilde{\chi}^-}
&=&
\frac{g_Z^2}{8 M_Z^2} C^{L\ast}_{d A X} C^{L}_{s B Y} 
\nonumber \\
&& 
\times 
\Bigg\{ 
       \delta_{XY} 
       \left[\sum_{j=1}^3 
               [U^\ast(\tilde{u})]_{jA} 
               [U(\tilde{u})]_{jB} 
            -\delta_{AB} 
             \left(1+2 T^3_{d_L}\right)
       \right]\, 
       I^1_3 ( m_{ \tilde{\chi}_X^-}^{2}, 
               m_{\tilde{u}_A}^{2},
               m_{\tilde{u}_B}^2)         
\nonumber \\
&& 
+ \delta_{AB} 
         \Bigg[ \,2  m_{ \tilde{\chi}_X^-} 
                  m_{ \tilde{\chi}_Y^-} 
                  [U^\ast(\tilde{\chi}^-_L)]_{1X} 
                  [U(\tilde{\chi}^-_L)]_{1Y} \, 
                  I^0_3( m_{ \tilde{\chi}_X^-}^{2}, 
                         m_{ \tilde{\chi}_Y^-}^{2},
                         m_{\tilde{u}_A}^2)       
\nonumber \\
&& 
         -  [U^\ast(\tilde{\chi}^-_R)]_{1X} 
                 [U(\tilde{\chi}^-_R)]_{1Y} \, 
                  I^1_3 ( m_{ \tilde{\chi}_X^-}^{2}, 
                          m_{ \tilde{\chi}_Y^-}^{2},
                          m_{\tilde{u}_A}^2)
\Bigg]                   
\Bigg\},
\\
\tilde{C}_{LL}^{ZP, \nu}|^{\tilde{\chi}^0}
&=&
\frac{g_Z^2}{8 M_Z^2} N^{L\ast}_{d A X} N^{L}_{s B Y} 
\nonumber \\
&& 
\times 
\Bigg\{ 
       \delta_{XY} 
       \left[-\sum_{j=1}^3 
               [U^\ast(\tilde{u})]_{jA} 
               [U(\tilde{u})]_{jB} 
            -2 T^3_{d_L}\delta_{AB} 
       \right]\, 
       I^1_3 ( m_{ \tilde{\chi}_X^0}^{2}, 
               m_{\tilde{d}_A}^{2},
               m_{\tilde{d}_B}^2)         
\nonumber \\
&&
+ \delta_{AB} 
         \left[\,-2  m_{ \tilde{\chi}_X^0} 
                  m_{ \tilde{\chi}_Y^0} 
                  O_L^{XY} \, 
                  I^0_3( m_{ \tilde{\chi}_X^-}^{2}, 
                         m_{ \tilde{\chi}_Y^-}^{2},
                         m_{\tilde{d}_A}^2)       
         +  O_R^{XY} \, I^1_3 ( m_{ \tilde{\chi}_X^0}^{2}, 
                          m_{ \tilde{\chi}_Y^0}^{2},
                          m_{\tilde{d}_A}^2)
          \right]                   
\Bigg\},
\nonumber \\
\end{eqnarray}
with
$g_Z=\sqrt{g_1^2+g_2^2}$ and 
\begin{eqnarray}
{H^L_{ud}}
&=&
\frac{g_2 \cot\beta\, m_u}{\sqrt{2} M_W} 
\left[V_{\rm KM}\right]_{ud},
\\
{H^R_{ud}}
&=&
\frac{g_2 \tan\beta\, m_d}{\sqrt{2} M_W} 
[V_{\rm KM}]_{ud}, 
\\
O_{L}^{XY}
&=&
[U^\ast(\tilde{\chi}^0)]_{3X}
[U(\tilde{\chi}^0)]_{3Y}
-
[U^\ast(\tilde{\chi}^0)]_{4X}
[U(\tilde{\chi}^0)]_{4Y},
\\
O_{R}^{XY}
&=&
-O_{L}^{\ast XY}.
\end{eqnarray}
Here $T^3_{u_R}=T^3_{d_R}=0,\ T^3_{u_L}=-T^3_{d_L}=1/2$.
$\tilde{C}_{RL}^{ZP,\nu}$ is obtained by interchanging $R$ with $L$,
$\tilde{C}_{RL}^{ZP,\nu}=\tilde{C}_{LL}^{ZP,\nu}|_{L\leftrightarrow R}$.

The contributions from box-diagrams are given by
\begin{eqnarray}
\tilde{C}_{LL}^{{\rm Box}, \nu}|^{\tilde{\chi}^-}
&=&
-\frac{1}{2} m_{ \tilde{\chi}_X^-} m_{ \tilde{\chi}_Y^-} 
C^{L\ast}_{dAX} C^{L}_{sAY}C^{L\ast}_{\nu BX} C^{L}_{\nu BY}
I^0_4 ( m_{ \tilde{\chi}_X^-}^{2}, 
        m_{ \tilde{\chi}_Y^-}^{2},
        m_{\tilde{u}_A}^2,
        m_{\tilde{l}_B}^2),
\\
\tilde{C}_{RL}^{{\rm Box},\nu}|^{\tilde{\chi}^-}
&=&
\frac{1}{4} 
C^{R\ast}_{dAX} C^{R}_{sAY}C^{L\ast}_{\nu BX} C^{L}_{\nu BY}
I^1_4 ( m_{ \tilde{\chi}_X^-}^{2}, 
        m_{ \tilde{\chi}_Y^-}^{2},
        m_{\tilde{u}_A}^2,
        m_{\tilde{l}_B}^2),
\\
\tilde{C}_{LL}^{{\rm Box}, \nu}|^{\tilde{\chi}^0}
&=&
-\frac{1}{4} 
N^{L\ast}_{\nu BX} N^{L}_{\nu BY}
\left[\,
       2 N^{L\ast}_{dAX} N^{L}_{sAY}
       m_{ \tilde{\chi}_X^0} m_{ \tilde{\chi}_Y^0} 
       I^0_4 ( m_{ \tilde{\chi}_X^0}^{2}, 
               m_{ \tilde{\chi}_Y^0}^{2},
               m_{\tilde{u}_A}^2,
               m_{\tilde{\nu}_B}^2)
\right.
\nonumber \\ &&
\left. 
+
       N^{L\ast}_{dAY} N^{L}_{sAX}
       I^1_4 ( m_{ \tilde{\chi}_X^0}^{2}, 
               m_{ \tilde{\chi}_Y^0}^{2},
               m_{\tilde{u}_A}^2,
               m_{\tilde{\nu}_B}^2)
\right],
\\
\tilde{C}_{RL}^{{\rm Box},\nu}|^{\tilde{\chi}^0}
&=&
\frac{1}{4} 
N^{L\ast}_{\nu BX} N^{L}_{\nu BY}
\left[\,
       2 N^{R\ast}_{dAY} N^{R}_{sAX}
       m_{ \tilde{\chi}_X^0} m_{ \tilde{\chi}_Y^0} 
       I^0_4 ( m_{ \tilde{\chi}_X^0}^{2}, 
               m_{ \tilde{\chi}_Y^0}^{2},
               m_{\tilde{u}_A}^2,
               m_{\tilde{\nu}_B}^2)
\right.
\nonumber \\ && 
\left. 
+
       N^{R\ast}_{dAX} N^{R}_{sAY}
       I^1_4 ( m_{ \tilde{\chi}_X^0}^{2}, 
               m_{ \tilde{\chi}_Y^0}^{2},
               m_{\tilde{u}_A}^2,
               m_{\tilde{\nu}_B}^2)
\right],
\end{eqnarray}
where $C^L_{\nu AX},\ N^L_{\nu AX}$ are vertices for 
$\tilde{\chi}^- - \nu - \tilde{l}$, $\tilde{\chi}^0 - \nu - \tilde{\nu}$,
respectively, that are analogue to those in the quark sector.

\section{SUSY Contribution to the $\Delta S=1$ Wilson Coefficients}
\label{app:dS=1}
\setcounter{equation}{0}

In this Appendix, we present the formulae for the SUSY contribution to
the Wilson coefficients of $\Delta S=1$ operators which are used for
the calculation of $\epsilon'/\epsilon$.  To make the notation
simpler, we first give the formulae for the Wilson coefficients for
the following operator basis:
\begin{eqnarray}
    {\cal H}_{\rm eff}^{(\Delta S=1)} &=& 
    (\bar{s}_\alpha \gamma_\mu P_L d_\alpha) 
    \sum_{q} \left[ 
        C_{LL}^{q({\rm S})} (\bar{q}_\beta \gamma_\mu P_L q_\beta)
        + C_{LR}^{q({\rm S})} (\bar{q}_\beta \gamma_\mu P_R q_\beta)
    \right]
    \nonumber \\ && +
    (\bar{s}_\alpha \gamma_\mu P_L d_\beta) 
    \sum_{q} \left[
        C_{LL}^{q({\rm T})} (\bar{q}_\beta \gamma_\mu P_L q_\alpha)
        + C_{LR}^{q({\rm T})} (\bar{q}_\beta \gamma_\mu P_R q_\alpha)
    \right]
    \nonumber \\ && +
    (L\leftrightarrow R).
\label{sd-qq}
\end{eqnarray}
The Wilson coefficients in the basis given in Eq.\ (\ref{sd-qq}) is
converted to the Wilson coefficients for the operators ${\cal
Q}_3^{(\Delta S=1)}$ $-$ ${\cal Q}_{10}^{(\Delta S=1)}$ as
\begin{eqnarray}
    C_{L,3}^{(\Delta S=1)} &=& 
    \frac{1}{12} C_{LL}^{u({\rm S})} 
    + \frac{1}{6} C_{LL}^{d({\rm S})},
    \\
    C_{L,4}^{(\Delta S=1)} &=& 
    \frac{1}{12} C_{LL}^{u({\rm T})} 
    + \frac{1}{6} C_{LL}^{d({\rm T})},
    \\
    C_{L,5}^{(\Delta S=1)} &=& 
    \frac{1}{12} C_{LR}^{u({\rm S})} 
    + \frac{1}{6} C_{LR}^{d({\rm S})},
    \\
    C_{L,6}^{(\Delta S=1)} &=& 
    \frac{1}{12} C_{LR}^{u({\rm T})} 
    + \frac{1}{6} C_{LR}^{d({\rm T})},
    \\
    C_{L,7}^{(\Delta S=1)} &=& 
    \frac{1}{6} C_{LR}^{u({\rm S})}
    - \frac{1}{6} C_{LR}^{d({\rm S})},
    \\
    C_{L,8}^{(\Delta S=1)} &=& 
    \frac{1}{6} C_{LR}^{u({\rm T})}
    - \frac{1}{6} C_{LR}^{d({\rm T})},
    \\
    C_{L,9}^{(\Delta S=1)} &=& 
    \frac{1}{6} C_{LL}^{u({\rm S})}
    - \frac{1}{6} C_{LL}^{d({\rm S})},
    \\
    C_{L,10}^{(\Delta S=1)} &=& 
    \frac{1}{6} C_{LL}^{u({\rm T})}
    - \frac{1}{6} C_{LL}^{d({\rm T})},
\end{eqnarray}
and $C_{R,i}^{(\Delta S=1)}=C_{L,i}^{(\Delta S=1)}|_{L\leftrightarrow
R}$.

Let us first consider the effect of gluon-penguin operator which
contributes only to the $\Delta I=\frac{1}{2}$ amplitude.  Denoting
\begin{eqnarray}
    \tilde{C}^{\rm GP}_L &=& 
    g_3 G^{L*}_{sA} G^{L}_{dA} \Bigg[
    \frac{1}{12N_c} I^2_5
    (m_{\tilde{G}}^2, m_{\tilde{d}_A}^2, m_{\tilde{d}_A}^2, 
    m_{\tilde{d}_A}^2, m_{\tilde{d}_A}^2)
    \nonumber \\ &&
    -\frac{N_c}{6} I^2_5
    (m_{\tilde{G}}^2, m_{\tilde{G}}^2, m_{\tilde{G}}^2
    , m_{\tilde{G}}^2, m_{\tilde{d}_A}^2)
    + \frac{N_c}{4} m_{\tilde{G}}^2 I^1_5
    (m_{\tilde{G}}^2, m_{\tilde{G}}^2, m_{\tilde{G}}^2
    , m_{\tilde{G}}^2, m_{\tilde{d}_A}^2)
    \Bigg]
    \nonumber \\ &&
    -\frac{1}{6} g_3 C^{L*}_{sAX} C^{L}_{dAX} I^2_5
    (m_{\tilde{\chi}^\pm_X}^2, m_{\tilde{u}_A}^2
    , m_{\tilde{u}_A}^2, m_{\tilde{u}_A}^2, m_{\tilde{u}_A}^2)
    \nonumber \\ &&
    -\frac{1}{6} g_3 N^{L*}_{sAX} N^{L}_{dAX} I^2_5
    (m_{\tilde{\chi}^0_X}^2, m_{\tilde{d}_A}^2
    , m_{\tilde{d}_A}^2, m_{\tilde{d}_A}^2, m_{\tilde{d}_A}^2),
\label{C(GP)}
\end{eqnarray}
we obtain
\begin{eqnarray}
    \left. \tilde{C}_{LL}^{q({\rm S})} \right|_{\rm GP} =
    \left. \tilde{C}_{LR}^{q({\rm S})} \right|_{\rm GP} =
    \frac{1}{2N_c} g_3 \tilde{C}^{\rm GP}_L,~~~
    \left. \tilde{C}_{LL}^{q({\rm T})} \right|_{\rm GP} =
    \left. \tilde{C}_{LR}^{q({\rm T})} \right|_{\rm GP} =
    - \frac{1}{2} g_3 \tilde{C}^{\rm GP}_L.
\end{eqnarray}
Notice that, in Eq.\ (\ref{C(GP)}) and hereafter, summation over the
dummy indices ($A$, $X$, and so on) is implied.

Contribution of the photon-penguin diagrams is parameterized by
\begin{eqnarray}
    \tilde{C}^{\rm PP}_L &=& 
    \frac{N_c^2-1}{36N_c} e G^{L*}_{sA} G^{L}_{dA} I^2_5
    (m_{\tilde{G}}^2, m_{\tilde{d}_A}^2, m_{\tilde{d}_A}^2, 
    m_{\tilde{d}_A}^2, m_{\tilde{d}_A}^2)
    \nonumber \\ &&
    + e C^{L*}_{sAX} C^{L}_{dAX} \Bigg[
    -\frac{1}{9} I^2_5
    (m_{\tilde{\chi}^\pm_X}^2, m_{\tilde{u}_A}^2, m_{\tilde{u}_A}^2, 
    m_{\tilde{u}_A}^2, m_{\tilde{u}_A}^2)
    \nonumber \\ &&
    +\frac{1}{3} I^2_5
    (m_{\tilde{\chi}^\pm_X}^2, m_{\tilde{\chi}^\pm_X}^2, 
    m_{\tilde{\chi}^\pm_X}^2, m_{\tilde{\chi}^\pm_X}^2, 
    m_{\tilde{u}_A}^2)
    \nonumber \\ &&
    -\frac{1}{2} m_{\tilde{\chi}^\pm_X}^2 I^1_5
    (m_{\tilde{\chi}^\pm_X}^2, m_{\tilde{\chi}^\pm_X}^2, 
    m_{\tilde{\chi}^\pm_X}^2, m_{\tilde{\chi}^\pm_X}^2, 
    m_{\tilde{u}_A}^2)
    \Bigg]
    \nonumber \\ &&
    + \frac{1}{18} e N^{L*}_{sAX} N^{L}_{dAX} I^2_5
    (m_{\tilde{\chi}^0_X}^2, m_{\tilde{d}_A}^2, m_{\tilde{d}_A}^2, 
    m_{\tilde{d}_A}^2, m_{\tilde{d}_A}^2).
\end{eqnarray}
With $\tilde{C}^{\rm PP}_L$, the photon-penguin contributions to the
Wilson coefficients are given by
\begin{eqnarray}
    \left. \tilde{C}_{LL}^{q({\rm S})} \right|_{\rm PP} =
    \left. \tilde{C}_{LR}^{q({\rm S})} \right|_{\rm PP} =
    - e e_q \tilde{C}^{\rm PP}_L,~~~
    \left. \tilde{C}_{LL}^{q({\rm T})} \right|_{\rm PP} =
    \left. \tilde{C}_{LR}^{q({\rm T})} \right|_{\rm PP} = 0.
\end{eqnarray}

Finally, we present the box contributions.  First, the box
contributions with internal gluino and/or neutralino lines are given
by
\begin{eqnarray}
    \left. \tilde{C}_{LL}^{q({\rm S})} 
    \right|_{\rm Box}^{(\tilde{G}, \tilde{\chi}^0)}
    &=&
    - G^{L*}_{sB} G^{L}_{dB} G^{L*}_{qB} G^{L}_{qB} 
    \Bigg[ \frac{1}{16N_c^2} I^1_4 
    (m_{\tilde{G}}^2, m_{\tilde{G}}^2, 
    m_{\tilde{d}_A}^2, m_{\tilde{q}_B}^2)
    \nonumber \\ &&
    + \frac{N_c^2+1}{8N_c^2} m_{\tilde{G}}^2 I^0_4 
    (m_{\tilde{G}}^2, m_{\tilde{G}}^2, 
    m_{\tilde{d}_A}^2, m_{\tilde{q}_B}^2)
    \Bigg]
    \nonumber \\ &&
    +\frac{1}{8N_c}
    (G^{L*}_{sA} N^{L}_{dAX} N^{L*}_{qBX} G^{L}_{qB}
    + N^{L*}_{sAX} G^{L}_{dA} G^{L*}_{qB} N^{L}_{qBX}) I^1_4 
    (m_{\tilde{G}}^2, m_{\tilde{\chi}^0_X}^2, 
    m_{\tilde{d}_A}^2, m_{\tilde{q}_B}^2)
    \nonumber \\ &&
    +\frac{1}{4N_c}
    (G^{L*}_{sA} N^{L}_{dAX} G^{L*}_{qB} N^{L}_{qBX}
    + N^{L*}_{sAX} G^{L}_{dA} N^{L*}_{qBX} G^{L}_{qB}) 
    \nonumber \\ && \times
    m_{\tilde{G}} m_{\tilde{\chi}^0_X} I^0_4 
    (m_{\tilde{G}}^2, m_{\tilde{\chi}^0_X}^2, 
    m_{\tilde{d}_A}^2, m_{\tilde{q}_B}^2)
    \nonumber \\ &&
    - \frac{1}{4} 
    N^{L*}_{sAX} N^{L}_{dAY} N^{L*}_{qBY} N^{L}_{qBX} I^1_4 
    (m_{\tilde{\chi}^0_X}^2, m_{\tilde{\chi}^0_Y}^2,
    m_{\tilde{d}_A}^2, m_{\tilde{q}_B}^2)
    \nonumber \\ &&
    - \frac{1}{2} 
    N^{L*}_{sAX} N^{L}_{dAY} N^{L*}_{qBX} N^{L}_{qBY} 
    m_{\tilde{\chi}^0_X} m_{\tilde{\chi}^0_Y} I^0_4 
    (m_{\tilde{\chi}^0_X}^2, m_{\tilde{\chi}^0_Y}^2,
    m_{\tilde{d}_A}^2, m_{\tilde{q}_B}^2),
    \\
    \left. \tilde{C}_{LL}^{q({\rm T})} 
    \right|_{\rm Box}^{(\tilde{G}, \tilde{\chi}^0)} &=&
    - G^{L*}_{sA} G^{L}_{dA} G^{L*}_{qB} G^{L}_{qB} 
    \Bigg[ \frac{N_c^2-2}{16N_c} I^1_4 
    (m_{\tilde{G}}^2, m_{\tilde{G}}^2, 
    m_{\tilde{d}_A}^2, m_{\tilde{q}_B}^2)
    \nonumber \\ &&
    - \frac{1}{4N_c} m_{\tilde{G}}^2 I^0_4 
    (m_{\tilde{G}}^2, m_{\tilde{G}}^2, 
    m_{\tilde{d}_A}^2, m_{\tilde{q}_B}^2)
    \Bigg]
    \nonumber \\ &&
    - \frac{1}{8}
    (G^{L*}_{sA} N^{L}_{dAX} N^{L*}_{qBX} G^{L}_{qB} 
    + N^{L*}_{sAX} G^{L}_{dA} G^{L*}_{qB} N^{L}_{qBX} ) I^1_4 
    (m_{\tilde{G}}^2, m_{\tilde{\chi}^0_X}^2, 
    m_{\tilde{d}_A}^2, m_{\tilde{d}_B}^2)
    \nonumber \\ &&
    - \frac{1}{4}
    (G^{L*}_{sA} N^{L}_{dAX} G^{L*}_{qB} N^{L}_{qBX}
    + N^{L*}_{sAX} G^{L}_{dA} N^{L*}_{qBX} G^{L}_{qB}) 
    \nonumber \\ && \times
    m_{\tilde{G}} m_{\tilde{\chi}^0_X} I^0_4 
    (m_{\tilde{G}}^2, m_{\tilde{\chi}^0_X}^2, 
    m_{\tilde{d}_A}^2, m_{\tilde{q}_B}^2),
    \\
    \left. \tilde{C}_{LR}^{q({\rm S})}
    \right|_{\rm Box}^{(\tilde{G}, \tilde{\chi}^0)} &=&
    - G^{L*}_{sA} G^{L}_{dA} G^{R*}_{qB} G^{R}_{qB}
    \Bigg[ -\frac{N_c^2+1}{16N_c} I^1_4 
    (m_{\tilde{G}}^2, m_{\tilde{G}}^2, 
    m_{\tilde{d}_A}^2, m_{\tilde{q}_B}^2)
    \nonumber \\ &&
    - \frac{1}{8N_c^2} m_{\tilde{G}}^2 I^0_4 
    (m_{\tilde{G}}^2, m_{\tilde{G}}^2, 
    m_{\tilde{d}_A}^2, m_{\tilde{q}_B}^2)
    \Bigg]
    \nonumber \\ &&
    - \frac{1}{8N_c}
    (G^{L*}_{sA} N^{L}_{dAX} G^{R*}_{qB} N^{R}_{qBX} 
    + N^{L*}_{sAX} G^{L}_{dA} N^{R*}_{qBX} G^{R}_{qB}) I^1_4 
    (m_{\tilde{G}}^2, m_{\tilde{\chi}^0_X}^2, 
    m_{\tilde{d}_A}^2, m_{\tilde{d}_B}^2)
    \nonumber \\ &&
    - \frac{1}{4N_c}
    (G^{L*}_{sA} N^{L}_{dAX} N^{R*}_{qBX} G^{R}_{qB} 
    + N^{L*}_{sAX} G^{L}_{dA} G^{R*}_{qB} N^{R}_{qBX}) 
    \nonumber \\ && \times
    m_{\tilde{G}} m_{\tilde{\chi}^0_X} I^0_4 
    (m_{\tilde{G}}^2, m_{\tilde{\chi}^0_X}^2, 
    m_{\tilde{d}_A}^2, m_{\tilde{q}_B}^2)
    \nonumber \\ &&
    + \frac{1}{4} 
    N^{L*}_{sAX} N^{L}_{dAY}  N^{R*}_{qBX} N^{R}_{qBY} I^1_4 
    (m_{\tilde{\chi}^0_X}^2, m_{\tilde{\chi}^0_Y}^2,
    m_{\tilde{d}_A}^2, m_{\tilde{q}_B}^2)
    \nonumber \\ &&
    + \frac{1}{2} 
    N^{L*}_{sAX} N^{L}_{dAY} N^{R*}_{qBY} N^{R}_{qBX} 
    m_{\tilde{\chi}^0_X} m_{\tilde{\chi}^0_Y} I^0_4 
    (m_{\tilde{\chi}^0_X}^2, m_{\tilde{\chi}^0_Y}^2,
    m_{\tilde{d}_A}^2, m_{\tilde{q}_B}^2),
    \\
    \left. \tilde{C}_{LR}^{q({\rm T})} 
    \right|_{\rm Box}^{(\tilde{G}, \tilde{\chi}^0)} &=&
    - G^{L*}_{sA} G^{L}_{dA} G^{L*}_{qB} G^{L}_{qB} 
    \Bigg[ \frac{1}{8N_c} I^1_4 
    (m_{\tilde{G}}^2, m_{\tilde{G}}^2, 
    m_{\tilde{d}_A}^2, m_{\tilde{d}_B}^2)
    \nonumber \\ &&
    - \frac{N_c^2-2}{8N_c} m_{\tilde{G}}^2 I^0_4 
    (m_{\tilde{G}}^2, m_{\tilde{G}}^2, 
    m_{\tilde{d}_A}^2, m_{\tilde{d}_B}^2)
    \Bigg]
    \nonumber \\ &&
    + \frac{1}{8}
    (G^{L*}_{sA} N^{L}_{dAX} G^{R*}_{qB} N^{R}_{qBX} 
    + N^{L*}_{sAX} G^{L}_{dA} N^{R*}_{qBX} G^{R}_{qB}) I^1_4 
    (m_{\tilde{G}}^2, m_{\tilde{\chi}^0_X}^2, 
    m_{\tilde{d}_A}^2, m_{\tilde{d}_B}^2)
    \nonumber \\ &&
    + \frac{1}{4}
    (N^{L*}_{sAX} G^{L}_{dA} G^{R*}_{qB} N^{R}_{qBX} 
    + G^{L*}_{sA} N^{L}_{dAX} N^{R*}_{qBX} G^{R}_{qB})
    \nonumber \\ && \times
    m_{\tilde{G}} m_{\tilde{\chi}^0_X} I^0_4 
    (m_{\tilde{G}}^2, m_{\tilde{\chi}^0_X}^2, 
    m_{\tilde{d}_A}^2, m_{\tilde{q}_B}^2).
\end{eqnarray}
In addition, contributions with internal chargino lines exist, which
are given by
\begin{eqnarray}
    \left. \tilde{C}_{LL}^{u({\rm S})} 
    \right|_{\rm Box}^{(\tilde{\chi}^\pm)}
    &=&
    - \frac{1}{2} C^{L*}_{sBX} C^{L}_{dBY} C^{L*}_{uBX} C^{L}_{uBY}
    m_{\tilde{\chi}^\pm_X} m_{\tilde{\chi}^\pm_Y} I^0_4 
    (m_{\tilde{\chi}^\pm_X}, m_{\tilde{\chi}^\pm_Y},
    m_{\tilde{d}_A}^2, m_{\tilde{q}_B}^2),
    \\
    \left. \tilde{C}_{LR}^{u({\rm S})} 
    \right|_{\rm Box}^{(\tilde{\chi}^\pm)}
    &=&
    \frac{1}{4} C^{L*}_{sBX} C^{L}_{dBY} C^{R*}_{uBX} C^{R}_{uBY}
    I^1_4 
    (m_{\tilde{\chi}^\pm_X}, m_{\tilde{\chi}^\pm_Y},
    m_{\tilde{d}_A}^2, m_{\tilde{d}_B}^2),
    \\
    \left. \tilde{C}_{LL}^{d({\rm S})} 
    \right|_{\rm Box}^{(\tilde{\chi}^\pm)}
    &=&
    - \frac{1}{4} C^{L*}_{sBX} C^{L}_{dBY} C^{R*}_{dBY} C^{R}_{dBX}
    I^1_4
    (m_{\tilde{\chi}^\pm_X}, m_{\tilde{\chi}^\pm_Y},
    m_{\tilde{u}_A}^2, m_{\tilde{u}_B}^2),
    \\
    \left. \tilde{C}_{LR}^{d({\rm S})} 
    \right|_{\rm Box}^{(\tilde{\chi}^\pm)}
    &=&
    \frac{1}{2} C^{L*}_{sBX} C^{L}_{dBY} C^{L*}_{dBY} C^{L}_{dBX}
    m_{\tilde{\chi}^\pm_X} m_{\tilde{\chi}^\pm_Y} I^0_4 
    (m_{\tilde{\chi}^\pm_X}, m_{\tilde{\chi}^\pm_Y},
    m_{\tilde{u}_A}^2, m_{\tilde{u}_B}^2),
\end{eqnarray}
while $\left.\tilde{C}_{LL}^{q({\rm T})}\right|_{\rm
Box}^{(\tilde{\chi}^\pm)}$ and $\left.\tilde{C}_{LR}^{q({\rm
T})}\right|_{\rm Box}^{(\tilde{\chi}^\pm)}$ vanish.

The Wilson coefficients $\tilde{C}^{q}_{RR}$ and $\tilde{C}^{q}_{RL}$
are obtained from $\tilde{C}^{q}_{LL}$ and $\tilde{C}^{q}_{LR}$ by
interchanging the indices $L$ and $R$.

\section{Master Integrals}
\label{app:I}
\setcounter{equation}{0}

In this Appendix, we present the master integrals used in the
calculations of the loop diagrams.

The function $I^N_D$ is defined as
\begin{eqnarray}
  I^N_D (m_1^2, m_2^2, \cdots, m_D^2) \equiv
  \int \frac{d^{4-2\epsilon}k}{(2\pi)^{4-2\epsilon}i}
  \frac{(k^2)^N}{(k^2-m_1^2)(k^2-m_2^2)\cdots (k^2-m_D^2)}.
\end{eqnarray}
Explicit formulae of $I^n_d$ used in our calculations are as follows:
\begin{eqnarray}
%
%
%
%
&& \hspace{-1.5cm} I^1_6 (M_X^2,M_X^2,M_X^2,M_X^2,m_A^2,m_A^2)
=
{\frac{1 + 9\,x - 9\,{x^2} - {x^3} + 6\,x\,\left( 1 + x \right) \,\log x}
   {48\,{{\pi }^2}\,{{M_X}^6}\,{{\left( x - 1 \right) }^5}}},
\hspace{2cm}\\
&& \hspace{-1.5cm} I^0_6 (M_X^2,M_X^2,M_X^2,M_X^2,m_A^2,m_A^2)
=
{\frac{17 - 9\,x - 9\,{x^2} + {x^3} + 6\,\left( 1 + 3\,x \right) \,\log x}
   {96\,{{\pi }^2}\,{{M_X}^8}\,{{\left( x - 1 \right) }^5}}},
\\
&& \hspace{-1.5cm} I^2_5 (M_X^2,m_A^2,m_A^2,m_A^2,m_A^2)
=
{\frac{11 - 18\,x + 9\,{x^2} - 2\,{x^3} + 6\,\log x}
   {96\,{{\pi }^2}\,{{M_X}^2}\,{{\left( x - 1 \right) }^4}}},
\\
&& \hspace{-1.5cm} I^1_5(M_X^2,m_A^2,m_A^2,m_A^2,m_A^2)
=
{\frac{2 + 3\,x - 6\,{x^2} + {x^3} + 6\,x\,\log x}
   {96\,{{\pi }^2}\,{{M_X}^4}\,{{\left( x - 1 \right) }^4}\,x}},
\\
&& \hspace{-1.5cm} I^1_4(M_X^2,M_X^2,m_A^2,m_A^2)
=
{\frac{1 - {x^2} + 2\,x\,\log x}
   {16\,{{\pi }^2\,{M_X^2}}\,{{\left( x - 1 \right) }^3}}},
\\
&& \hspace{-1.5cm} I^0_4(M_X^2,M_X^2,m_A^2,m_A^2)
=
{\frac{2 - 2\,x + \left( 1 + x \right) \,\log x}
   {16\,{{\pi }^2\,{{{M_X}}^4}}\,{{\left( x - 1 \right) }^3}}},
\\
&& \hspace{-1.5cm} I^1_3 (M_X^2,M_X^2,m_A^2)
=
\frac{x(-1+x- x\,\log x) 
     }{ 16\,\pi^2 ( x - 1)}
+I_{\rm Div},
\\
&& \hspace{-1.5cm} I^0_3 (M_X^2,m_A^2,m_A^2)
=
{\frac{1 - x + \log x}
   {16\,{{\pi }^2\,{{{M_X}}^2}}\,{{\left( x - 1 \right) }^2}}},
\\
&& \hspace{-1.5cm} I^0_2 (M_X^2,m_A^2)
=
{\frac{ - 1 + x  - x\,\log x}
   {16\,{{\pi }^2}\,\left( x - 1 \right) }}
+I_{\rm Div},
\\
&& \hspace{-1.5cm} I^1_4(M_X^2,M_X^2,m_A^2,m_B^2)
=
{\frac{1}{16\,{{\pi }^2\,{{{M_X}}^2}}\,\left( x - y \right) }}
\left[ {\frac{-1+x-{x^2}\,\log x}{{{\left( x - 1 \right) }^2}}}
-(x \rightarrow y) \right],
\hspace{1.5cm} \\
&& \hspace{-1.5cm} I^0_4(M_X^2,M_X^2,m_A^2,m_B^2)
=
\frac{1}{16\,{{\pi }^2\,{M_X^4}}\,(x-y)}
\left[ \frac{-1+x - {x} \, \log x}{ \left( x - 1 \right)^2} 
- (x \rightarrow y)
\right],
\\
&& \hspace{-1.5cm} I^1_3 (M_X^2,m_A^2,m_B^2)
=
\frac{1}{16\,\pi^2 ( x - y) } 
\left[
      \frac{-x+x^2-x^2 \,\log x}{x - 1}
    - (x \rightarrow y)
\right]
+I_{\rm Div},
\\ 
&& \hspace{-1.5cm} I^0_3 (M_X^2,m_A^2,m_B^2)
=
\frac{1}{16\,\pi^2\,{M_X}^2  \,\left( x - y \right)}
\left[\frac{-x\,\log x}{ x -1 }-(x\rightarrow y) 
\right],
\\
&& \hspace{-1.5cm} I^1_4 (M_X^2,m_A^2,m_B^2,m_C^2)
=
\frac{1}{16\,{{\pi }^2}\,{M_X^2}\,(x-y)}
\left\{ \frac{1}{x-z} \left[
              \frac{- {x^2} \, \log x }{ \left( x - 1 \right)} 
               -(x \rightarrow z)
                    \right]
- (x \rightarrow y)
\right\},
\nonumber \\ \\
&& \hspace{-1.5cm} I^0_4(M_X^2,m_A^2,m_B^2,m_C^2)
=
\frac{1}{16\,{{\pi }^2\,{M_X^4}}\,(x-y)}
\left\{ \frac{1}{x-z} \left[
              \frac{- {x} \, \log x }{ \left( x - 1 \right)} 
               -(x \rightarrow z)
                    \right]
- (x \rightarrow y)
\right\},
\nonumber \\
%
%
%
\end{eqnarray}
where $x=m_A^2/M_X^2,\, y=m_B^2/M_X^2,\, z=m_C^2/M_X^2$, and $I_{\rm
  Div}$ contains divergence of $1/\epsilon$ in dimensional
regularization:
\begin{eqnarray}
I_{\rm Div}= \frac{1}{16\, \pi^2}
\left[\frac{1}{\epsilon}-\gamma_E+\log 
\left( \frac{4 \pi}{M_X^2} \right) \right].
\end{eqnarray}


\begin{thebibliography}{99}

\bibitem{BELLE}
    BELLE Collaboration (A. Abashian et al.),
    {\sl Phys.\ Rev.\ Lett.} {\bf 86} (2001) 2509.

\bibitem{BABAR}
    BABAR Collaboration (B. Aubert et al.),
    {\sl Phys.\ Rev.\ Lett.} {\bf 86} (2001) 2515.

\bibitem{PTP49-652}
    M. Kobayashi and T. Maskawa,
    {\sl Prog.\ Theor.\ Phys.} {\bf 49} (1973) 652.

\bibitem{wolfenstein51}
    L. Wolfenstein,
    {\sl Phys.\ Rev.\ Lett.} {\bf 51} (1983) 1945.

\bibitem{LHCB}
    F. Muheim (LHCb Collaboration),
    hep-ex/0012059.
    
\bibitem{KOPIO}
    KOPIO Collaboration,
    {\tt http://pubweb.bnl.gov/people/rsvp/proporsal.ps}.

\bibitem{SUPERK}
    Super-Kamiokande Collaboration (Y. Fukuda et al.),
    {\sl Phys.\ Rev.\ Lett.} {\bf 81} (1998) 1562.

\bibitem{SOLAR}
    Super-Kamiokande Collaboration (S. Fukuda et al.),
    hep-ex/0103033.

\bibitem{seesaw}
    T.\ Yanagida, 
    in {\sl Proceedings of the Workshop on Unified Theory and Baryon
    Number of the Universe}, eds.\ O.\ Sawada and A.\ Sugamoto (KEK,
    1979) p.95;\\
    M.\ Gell-Mann, P.\ Ramond and R.\ Slansky, 
    in {\sl Supergravity}, eds.\ P.\ van Niewwenhuizen and D.\ 
    Freedman (North Holland, Amsterdam, 1979).

\bibitem{moroi00}
    T. Moroi 
    {\sl JHEP} {\bf 0003} (2000) 019; \\
    T. Moroi 
    {\sl Phys.\ Lett.\ B} {\bf 493} (2000) 366.

\bibitem{baek00}
    S. Baek, T. Goto, Y. Okada, K.-I. Okumura,
    {\sl Phys.\ Rev.} {\bf D63} (2001) 051701.

\bibitem{hall86}
    L.J. Hall, V.A. Kostelecky and S. Raby, 
    {\sl Nucl.\ Phys.} {\bf B267} (1986) 415.

\bibitem{kurimoto89}
    T.\ Kurimoto,
    {\sl Phys.\ Rev.} {\bf D39} (1989) 3447.

\bibitem{bertolini91}
    S. Bertolini, F. Borzumati, A. Masiero and G. Ridolfi,
    {\sl Nucl.\ Phys.} {\bf B353} (1991) 591.

\bibitem{gotookada}
    T.\ Goto, Y.\ Okada and T.\ Nihei,
    {\sl Phys.\ Rev.} {\bf D53} (1996) 5233, 
    Erratum {\sl ibid.} {\bf D54} (1996) 5904;\\
    T.\ Goto, Y.\ Okada and Y.\ Shimizu,
    hep-ph/9908449;\\
    T. Goto, Y. Okada and Y. Shimizu,
    {\sl Rhys.\ Rev.} {\bf D58} (1998) 094006.

\bibitem{maki62}
    Z. Maki, M. Nakagawa, S. Sakata, 
    {\sl Prog.\ Theor.\ Phys.} {\bf 28} (1962) 870.

\bibitem{PDG}
    D.E. Groom et al,
    {\sl Eur.\ Phys.\ J.} {\bf C15} (2000) 1.

\bibitem{arkanihamed96}
    N.\ Arkani-Hamed, H.C.\ Cheng and L.J.\ Hall,
    {\sl Phys.\ Rev.} {\bf D53} (1996) 413.

\bibitem{ciafaloni96}
    P.\ Ciafaloni, A.\ Romanino and A.\ Strumia,
    {\sl Nucl.\ Phys.} {\bf B458} (1996) 3.

\bibitem{HNOST98}
    J.\ Hisano, D.\ Nomura, Y.\ Okada, Y.\ Shimizu and M.\ Tanaka,
    {\sl Phys.\ Rev.} {\bf D58} (1998) 116010.

\bibitem{BilenkyGiunti}
    S.M.\ Bilenky and C.\ Giunti, hep-ph/9802201.

\bibitem{CHOOZ}
    CHOOZ Collaboration (M. Apollonio et al.),
    {\sl Phys.\ Lett.} {\bf B466} (1999) 415. 

\bibitem{borzumati86}
    F.~Borzumati and A.~Masiero, 
    {\sl Phys.\ Rev.\ Lett.} {\bf 57} (1986) 961.

\bibitem{HMTYNR}
    J.\ Hisano, T.\ Moroi, K.\ Tobe, M.\ Yamaguchi and T.\ Yanagida,
    {\sl Phys.\ Lett.} {\bf B357} (1995) 579;\\
    J.\ Hisano, T.\ Moroi, K.\ Tobe and M.\ Yamaguchi,
    {\sl Phys.\ Rev.} {\bf D53} (1996) 2442.

\bibitem{HisanoNomura}
    J.\ Hisano and D.\ Nomura,
    {\sl Phys.\ Rev.} {\bf D59} (1999) 116005.

\bibitem{barbierihall}
    R.~Barbieri and L.J.~Hall,
    {\sl Phys.\ Lett.} {\bf B338} (1994) 212;\\
    R.\ Barbieri, L.J.\ Hall and A.\ Strumia,
    {\sl Nucl.\ Phys.} {\bf B445} (1995) 219.

\bibitem{HMTYSU5}
    J.\ Hisano, T.\ Moroi, K.\ Tobe and M.\ Yamaguchi,
    {\sl Phys.\ Lett.} {\bf B391} (1997) 341

\bibitem{BarbieriHallStrumia}
    R.\ Barbieri, L.J.\ Hall and A.\ Strumia,
    {\sl Nucl.\ Phys.} {\bf B449} (1995) 437.

\bibitem{Matchev97}
    J. A. Bagger, K. T. Mathcv and R.-J. Zhang,
    {\sl Phys.\ Lett.} {\bf B412} (1997) 77.

\bibitem{MassInsertion}
    J.S. Hagelin, S. Kelley and T. Tanaka,
    {\sl Nucl.\ Phys.} {\bf B415} (1994) 293;\\
    F.\ Gabbiani, E.\ Gabrielli, A.\ Masiero and L.\ Silvestrini,
    {\sl Nucl.\ Phys.} {\bf B477} (1996) 321.

\bibitem{hph9704376}
    A.J. Buras and R. Fleischer,
    hep-ph/9704376.

\bibitem{inami81}
    T. Inami and C.S. Lim,
    {\sl Prog.\ Theor.\ Phys.} {\bf 65} (1981) 297.

\bibitem{buras96}
    G. Buchalla, A.J. Buras and M.E. Lautenbacher 
    {\sl Rev.\ Mod.\ Phys.} {\bf 68} (1996) 1125

\bibitem{PRL83-4929}
    A.L. Kagan and M. Neubert,
    {\sl Phys.\ Rev.\ Lett.} {\bf 83} (1999) 4929.

  \bibitem{KTeVNA48} KTeV Collaboration, {\sl Phys. Rev. Lett.} {\bf
      83} (1999) 22; \\ NA48 Collaboration, {\sl Phys. Lett.} {\bf
      B465} (1999) 335; {\tt http://www.cern.ch/NA48/Welcome.html}.

\bibitem{PRL83-907}
    A. Masiero and H. Murayama,
    {\sl Phys.\ Rev.\ Lett.} {\bf 83} (1999) 907.

\bibitem{baek01}
    S. Baek, T. Goto, Y. Okada and K.-I. Okamura,
    hep-ph/0104146.    

\end{thebibliography}
\end{document}